\documentclass[11pt]{frank} 

\usepackage{amsmath, amscd, amsthm, amsfonts, amssymb}
\usepackage{mathrsfs, bbm, cancel, stmaryrd, array}
\usepackage{comment}
\usepackage[FIGTOPCAP]{subfigure}
\usepackage[all]{xy}
\usepackage{graphicx,enumitem}
\usepackage{scrtime} 
\usepackage{rotating, lscape, wrapfig, tabularx, caption, color, microtype}
\usepackage[hyperfootnotes=false]{hyperref}
\usepackage{marvosym} 
\usepackage{color}
\usepackage{media9}
\definecolor{darkblue}{rgb}{0,0,1}
\hypersetup{pdfpagemode=UseNone, colorlinks=true, breaklinks=true, linkcolor=blue, menucolor=black, urlcolor=blue, citecolor=red}
\usepackage[bottom]{footmisc}
\usepackage{datetime}

\usepackage{nccfoots}
\usepackage{framed}

\newtheorem{satz}{Satz}

\usepackage{setspace}

\theoremstyle{remark}

\theoremstyle{definition}

\newtheorem{rem}[satz]{Remark}

\usepackage{titletoc}

\allowdisplaybreaks

\begin{document}
\sloppy \raggedbottom

\sloppy \raggedbottom

\title{On the approximation of D.I.Y.\ water rocket dynamics
including air drag}

\begin{start}

	\author{
			L.~Fischer\,}{1}{\bf }
	\author{
			T.~G\"unther\,}{1,a}{\bf }
	\author{
			L.~Herzig\,}{1}{\bf }
	\author{
			T.~Jarzina\,}{1}{\bf }
	\author{
			F.~Klinker\,}{2,3,b}{\bf }

	\author{
			S.~Knipper\,}{1}{\bf }
	\author{
			F.-G.~Sch\"urmann\,}{1}
	\author{
			M.~Wollek\,}{1}\\

	\address{Marie-Curie-Gymnasium, Billy-Montigny-Platz 5, 59199 B\"onen, Germany}{1}

	\address{TU Dortmund, Fakult\"at f\"ur Mathematik, 44221 Dortmund, Germany}{2}

	\address{Eduard-Spranger-Berufskolleg, Vorheider Weg 8, 59067 Hamm, Germany}{3}

\noindent
\makebox[0.8\textwidth]{%
\begin{minipage}{0.85\textwidth}
\begin{Abstract}
	If you want to get accurate predictions for the motion of water and air propelled D.I.Y rockets, neglecting
air resistance is not an option. 
But the theoretical analysis including air drag leads to a system of differential
equations which can only be solved numerically. 
We propose an approximation which simply works by the
estimate of a definite integral and which is even feasible for undergraduate physics courses. 
The results
only slightly deviate from the reference data (received by the Runge-Kutta method). 
The motion is divided
into several flight phases that are discussed separately and the resulting equations are solved by analytic and
numeric methods. The different results from the flight phases are collected and are compared to data that
has been achieved by well explained and documented experiments. 
Furthermore, we theoretically estimate
the rocket's drag coefficient. The result is confirmed by a wind tunnel experiment.
	\end{Abstract}
\end{minipage}}

\let\thefootnote\relax\footnote{ \ \\[-2ex]
${}^a$\,\href{mailto:dr.thomas.guenther@gmail.com}{dr.thomas.guenther@gmail.com} (corresponding author)\\
${}^b$\,\href{mailto:frank.klinker@math.tu-dortmund.de}{frank.klinker@math.tu-dortmund.de}\\[0.5ex]
{\em Int.~J.~of Sci.~Research in Mathematical and Statistical Sciences} {\bf 6} (2019) no.~6, 1-13\\[0.5ex]
We suggest the reader to choose this preprint version because the journal text contains several formatting errors.
} 

\renewcommand{\dateseparator}{-}
\end{start}

\runningheads{%
\small\sf Preprint 
\hspace*{4.9cm}
\sf On the approximation of  D.I.Y.\ water rocket dynamics
}{%
\small\sf Preprint 
\hspace*{10.4cm}   
\sf L.~Fischer et.~al.
}

\section{Introduction}

The classical rocket equation \cite[eq.\ 2.12]{Raumfahrtsysteme} is
attributed to Konstantin Eduardovich Tsiolkovsky \cite{Tsiolkovsky}.
If the propellant is exhausted with constant speed the rocket equation
can easily be integrated. In case of water and air propelled rockets
the exhaust speed decreases together with the internal pressure and
mass of the rocket. Physics of water rockets have been the subject
of several investigations. In an early paper \cite[p.\ 152]{Nelson1976},
Nelson and Wilson claim that rocket thrust and mass as a function
of time, as well as the drag coefficient must be determined experimentally.
Later, Finney assumes in \cite{Finney1999} that ``\emph{the air
in the rocket behaves as an ideal gas and} {[}...{]}\emph{ expands
isothermally}''. He uses Bernoulli's equation to determine
an equation for the propellant's mass flow rate, see \cite[eq.\ 3]{Finney1999}.
Furthermore, Finney deduces an equation for the pressure as a function
of time and calculated the rocket's burn time, see \cite[eq.\ 7, 8]{Finney1999}.
Therewith, he proposes the estimate
\begin{equation}
h=\frac{1}{8}\,g\,t^{2}-\frac{D}{64\,M}g^{2}t^{4}\label{h-Finney}
\end{equation}
for the rocket's height, where
$D=\frac{1}{2}\rho_{\text{air}}c_{D}A_{R}$ contains the parameters
of the drag force $F_{D}$ and $M$ is the mass of the empty rocket, see \cite[eq.\ 14]{Finney1999}.  
As mentioned
in the appendix of his paper ``\emph{an adiabatic approximation would
seem more natural}.'' Gommes slightly improves the thrust prediction
for water rockets: ``\emph{the gas expansion has to be modeled as
an adiabatic process.}'', see \cite{Gommes2010}. He also includes
the fact, that ``\emph{Air expansion }{[}...{]}\emph{ is accompanied
by vapor condensation}''. A more meticulous investigation of the
thermodynamics of the water rocket's thrust phase was published by
Romanelli, Bove and Madina in \cite{Air-Expansion-2013}. Indeed,
the value of the polytropic exponent $n$ in $pV^{n}=\text{constant}$
affects the rocket's maximum altitude. There are several other effects
that raise the inaccuracy of the predictions more than that. Prusa
used $n=1.4$ for dry air, see \cite[p.\ 724]{Prusa2000}. In our paper
we will take into account the enhanced value for $n$. Since $n$
is a constant, this doesn't cause additional difficulties. Now let
us get back to the rocket's movement: Prusa derived an equation for
the rocket's acceleration, cf.\ \cite[eq.\ 2.2]{Prusa2000}, and
proposed a numerical algorithm to solve it, see \cite[p.\ 723]{Prusa2000}.
Prusa neglected  additional thrust from  compressed air which
is left at the end of the water propulsion phase. An even more accurate
consideration of the water rocket physics was given by Barrio-Perott
et.\ al.\ in 2010, see \cite{Barrio2010}. They worked out a set of
differential equations for the water thrust, cf.\ \cite[eq.\ 18-21]{Barrio2010},
as well as equations for the air thrust \cite[eq.\ 28-31]{Barrio2010}.
Both articles, \cite{Prusa2000} and \cite{Barrio2010} use numerical
methods to achieve the solution. 

In our paper the rocket's ascent is divided into water thrust
phase, air thrust phase and upward coasting phase. We also set up the
equations of motion for the rocket in Section \ref{sec:AccelerationEquation-for-the}.
The equation that governs the gas expansion inside the rocket tank
is solved analytically in Subsection \ref{subsec:Analytic-solution-of},
but it is not possible to analytically solve the complete system of
differential equations that describe the rocket's launch. First and
foremost, we study the water thrust phase in Section \ref{sec:Movement-in-thrust}.
On the one hand, we apply the Runge-Kutta method to the corresponding
initial value problem to have some proper reference data. On the other
hand, we deduce a simple method to approximate the results with high
accuracy: Our approximation simply works by the calculation of a definite
integral. One advantage is that this calculation can be done with
a simple graphing calculator without a computer algebra system and
actually is feasible for undergraduate physics courses. Nonetheless,
one receives very accurate results which only slightly deviate from
the reference data (received by the Runge-Kutta method). Section \ref{sec:The-air-thrust}
is concerned with the air thrust phase. D.I.Y.\ rockets are chaotic
systems and there is no way to receive analytic results. Therefore,
we make reasonable assumptions to simplify the underlying equations
and introduce an efficiency factor for the air thrust. In Section
\ref{sec:Upward-Coast} the upward coasting phase is discussed. A collection
of the previous results is given in Section \ref{sec:Collection-of-theResults}.
Subsequently, our paper contains a detailed theoretical discussion
of the drag coefficient and we verify the results by experimental
data from a wind tunnel experiment, see Section \ref{sec:Drag-analysis}.
Finally, we compare our formula of the rocket's maximum altitude
with the experimental data from our D.I.Y.\ rocket launch experiments. 

\medskip
{\bf Acknowledgments.} We are grateful to Prof.\ Dr.\ Andreas Br\"ummer of
the Department of Fluidics at TU Dortmund University for kindly
providing us access to the department's wind tunnel.

\section{\label{sec:AccelerationEquation-for-the}The rocket's acceleration and exhaust velocity}

\subsection{The acceleration}

Consider a rocket moving with velocity $v$ in vertical direction.
Let $m$ be the mass of the rocket including its propellant at a given
time $t$. During the interval of time $dt$, the rocket ejects the
mass element $dm$ with the exhaust speed $v_{e}$. According to the
law of conservation of momentum, the velocity $v$ of the rocket thereby
increases by $dv=-v_{e}m^{-1}dm$, see e.g.\ \cite[eq.\ 2.11]{Raumfahrtsysteme}.
Consequently, the thrust acceleration is given by 
\begin{equation}
a_{I}=-\frac{dm}{dt}\cdot\frac{v_{e}}{m}.\label{a-thrust}
\end{equation}
Since the total mass is shrinking, i.e.\ $\frac{dm}{dt}<0$, it is
$a_{I}>0$. A rocket within the terrestrial atmosphere additionally
experiences deceleration from gravity and air drag. Both of them decrease
with the rocket's height. However, a simple water rocket does
only reach a maximum altitude of a few meters\footnote{It should be mentioned that a group of scientists at the University
of Cape Town built a water and air propelled rocket that reached 830\,$m$
in 2015. This is the current record, cf.\ \href{http://www.wra2.org/}{http://www.wra2.org/}
and \href{https://www.news.uct.ac.za/article/-2015-10-07-uct-team-smashes-eight-year-water-rocket-world-altitude-record}{https://www.news.uct.ac.za/article/-2015-10-07-uct-team-smashes-eight-year-water-rocket-world-altitude-record}.
Certainly, a simple home-made rocket built from a 1 or 2 liter PET
bottle is not even in a position to achieve this altitude.}. Consequently, the altitude dependence of gravity and air drag can
be neglected. For the magnitude $g=GM_{\text{earth}}r^{-2}$ of barycentric
gravitational acceleration we use $g=9.81{m}s^{-2}$ during
 numerical calculations. Air resistance is modeled by the drag force
$F_{D}=\frac{1}{2}\rho_{\text{air}}c_{D}A_{R}v^{2}$ where $\rho_{\text{air}}$
is the density of air, $c_{D}$ the drag coefficient, $A_{R}$
the reference area (later we use the rocket's cross sectional area $A_{cs}$
 perpendicular to the direction of movement), and $v$
 the rocket's velocity. Let us adopt the abbreviation 
\begin{equation}
D:=\frac{1}{2}\rho_{\text{air}}c_{D}A_{R}\label{D-def}
\end{equation}
from \cite{Finney1999}. The drag force $F_{D}=Dv^{2}$ leads to a
deceleration
\begin{equation}
a_{D}=\frac{F_{D}}{m}=\frac{D}{m}v^{2}.\label{aD-air-drag}
\end{equation}
An estimate of the drag coefficient of our model rocket is given
in Section \ref{sec:Drag-analysis}. Deceleration from gravity and
drag point on the one hand and thrust acceleration on the other hand point in opposite directions. 
Therefore,
the rocket's total acceleration  is given by 
\begin{equation}
a=a_{I}-g-a_{D}=-\frac{v_{e}dm}{m\,dt}-g-\frac{D}{m}v^{2}.\label{a_rakete-gesamt}
\end{equation}
For a water rocket, the working mass (water) represents the major
part of the rocket's total mass $m$. Anyway, the total mass strongly
decreases with time. In order to integrate \eqref{a_rakete-gesamt}
it is necessary to know the dependence of mass $m$ and time $t$. 

\subsection{Mass and exhaust velocity}

Let $\widetilde{m}\left(t\right)$ be the time dependent working mass,
namely the mass water in the rocket tank. Further let $M$
denote the constant mass of the empty rocket. The time dependent mass
of the compressed air doesn't contribute significantly to the total
mass of the rocket. Despite this fact, we will take it into account.
Section \eqref{sec:The-air-thrust} is concerned with the additional
air propulsion after the water thrust phase. During the water thrust
phase we treat the mass of compressed air as a constant, say $m_{\text{air}}$.
This approach is based on the idea that all  water is expelled before
the air escapes. Of course, this is somehow physically unrealistic.
Indeed, an mixture of water and air is expelled, which yields a non-calculable chaos, in particular towards
the end of the water thrust phase.
Within our model  the rocket's mass is given by $m(t)=M+m_{\text{air}}+\widetilde{m}(t)$
during the water thrust phase. Let $\rho$ denote the density of the
working mass, in case of water this is $\rho\approx1000\,{kg}\,m^{-3}$,
and $\widetilde{V}(t)$ its volume. $V_{b}$ is the volume
of the rocket's tank and $V(t)=V_{b}-\widetilde{V}(t)$
the part which is filled with air (or another gas, maybe water vapor
etc.). Therefore, the mass $m$ and the differential
$dm$ read
\begin{equation}
m=M+m_{\text{air}}+\rho\left(V_{b}-V(t)\right)\quad\Rightarrow\quad dm=-\rho\,dV.\label{mass-mt}
\end{equation}
Thereby, the dependence of mass $m$ and time $t$ is determined by
the gas expansion during the water thrust phase. Let $p_{a}$
denote the atmospheric pressure and $p=p\left(V\right)$ the pressure
of propellant and gas inside the rocket. Initial values at rocket
launch for gas pressure and volume are denoted by $p_{0}$ and $V_{0}$,
respectively. Bernoulli's equation $p=p_{a}+v_{e}^{2}\rho/2$, see
\cite{St=0000F6cker}, relates the exhaust speed $v_{e}$ to the pressure
difference $p-p_{a}$: 
\begin{equation}
v_{e}=\sqrt{\frac{2\left(p\left(V\right)-p_{a}\right)}{\rho}}.\label{ve}
\end{equation}
Indeed, the barometric pressure $p_{a}$ slightly depends on the altitude,
see \cite{Manual-Atmosphere}. At altitudes that can be reached by
a water rocket, $p_{a}$ decreases by about $12\,{Pa/m}=12^{-5}\,bar/m$.
Since the pressure of the propellant will usually be about a few bar we can neglect the altitude
dependence of the barometric pressure. Another point is that the fluid's
velocity before leaving the nozzle is assumed to be zero in \eqref{ve}.
Strictly speaking, we have to take into account the rejuvenation of
the bottle. Consider an incompressible fluid (like water) which laminar
flows from a point inside the bottle with cross-section $A_{R}$ through
the nozzle with cross-section $A$ at velocity $w_{e}$. Let $\widetilde{w}$
denote the velocity of the fluid inside the bottle. The continuity
equation leads to $\widetilde{w}=w_{e}A/A_{R}=w_{e}r^{2}/R^{2}$ where
$R$ and $r$ are the radii of bottle and nozzle, respectively. Usually,
the radius of the nozzle will be small against the rocket's radius,
that is $r\ll R$. From Bernoulli's equation $p+\left(w_{e}r^{2}/R^{2}\right)^{2}\rho/2=p_{a}+w_{e}^{2}\rho/2$
we receive the slightly more accurate exhaust velocity 
\[
w_{e}=\sqrt{\frac{2\left(p\left(V\right)-p_{a}\right)}{\rho}}\cdot\frac{1}{\sqrt{1-\left(\frac{r}{R}\right)^{4}}}\;\overset{\eqref{ve}}{=}\;v_{e}\left\{ 1+\frac{r^{4}}{2R^{4}}+\mathcal{O}\left(\left[\frac{r}{R}\right]^{8}\right)\right\} .
\]
As shown above\footnote{The corresponding Taylor series is ${\displaystyle \frac{1}{\sqrt{1-x^{4}}}=1+\frac{1}{2}x^{4}+\mathcal{O}\left(x^{8}\right)}$
with $x=r/R$.}, equation \eqref{ve} represents the third order Taylor approximation
w.\ r.\ t.\ $r/R$ for the exhaust velocity $w_{e}$. But what is the
 error while using \eqref{ve} instead of the latter
equation? The radius of a commercially available PET bottle can be
estimated to about $4$ to $5\,{cm}$. Our nozzle has a diameter
of about $9\,{mm}$. That leads to $R\approx10r$ and we get
$w_{e}\approx1.00005\,v_{e}$. The error for the exhaust velocity
is about $0.005\%$. Even for a nozzle with $r=R/3$, which
seems to be unrealistic large, the error will be less than 1\%.
Obviously, approximation \eqref{ve} is good enough for our concern.
The working mass pressure and therewith the pressure difference depend
on the gas volume $V$. The propellant is expelled through a nozzle
with a cross sectional area $A$ at speed $v_{e}$. Let $d\vec{A}$
be the vector surface element normal to $A$. In our case, the velocity
vector $\vec{v}_{e}$ is perpendicular to $A$ as well. Therefore,
the volumetric flow rate through the plane surface $A$ reduces to
\begin{equation}
\frac{dV}{dt}=\iint_{A}\left\langle \vec{v}_{e},d\vec{A}\right\rangle =Av_{e}\label{volumetric-flow-rate}
\end{equation}
and from \eqref{mass-mt} we deduce the mass flow rate
\begin{equation}
\frac{dm}{dt}=\frac{d}{dt}\left[M+m_{\text{air}}+\rho\left(V_{b}-V\left(t\right)\right)\right]=-\rho\frac{dV}{dt}\;\overset{\eqref{volumetric-flow-rate}}{=}\;-\rho Av_{e}.\label{mass-flow-rate}
\end{equation}
From \eqref{ve} one gets $\rho v_{e}^{2}=2\left(p\left(V\right)-p_{a}\right)$.
With the above considerations the rocket's acceleration \eqref{a_rakete-gesamt}
takes the form 
\begin{equation}
a=\frac{\rho Av_{e}^{2}}{m\,}-g-\frac{D}{m}v^{2}=\frac{2A\left(p\left(V\right)-p_{a}\right)-Dv^{2}}{M+m_{\text{air}}+\rho\left(V_{b}-V\right)}-g\label{a-Rakete-Vt}
\end{equation}
where $V$ is the gas volume as a function of time $t$. In order
to solve the latter equation it is also necessary to know the relation
$p\left(V\right)$ of pressure and volume.

\section{\label{subsec:Polytropic-expansion-during}Gas Expansion gas during
the thrust phase}

``\emph{Many real processes undergone by gases or vapours are approximately
polytropic with a polytropic index typically between 1.0 and 1.7}
{[}...{]}'', cf.\ \cite{Clifford2006}. It is argued in \cite{Gommes2010}
that the gas expansion in a model rocket can also be described by a
polytropic process. Pressure $p$ and Volume $V$ are related by $pV^{n}=constant$.
If $p_{0}$ and $V_{0}$ denote the corresponding initial values this
is
\begin{equation}
p\left(V\right)=p_{0}\left(\frac{V_{0}}{V}\right)^{n}.\label{p-polytrop}
\end{equation}
Volumetric flow rate \eqref{volumetric-flow-rate} and exhaust speed
\eqref{ve} lead to 
\begin{equation}
\frac{dV}{dt}=A\,v_{e}=A\sqrt{\frac{2\left(p\left(V\right)-p_{a}\right)}{\rho}}\label{dV-nach-dt}
\end{equation}
which together with \eqref{p-polytrop} yields 
\begin{equation}
\frac{dV}{dt}=A\sqrt{\frac{2\left(p_{0}V_{0}^{n}V^{-n}-p_{a}\right)}{\rho}}.\label{dVdt-mit-Polytrop-n}
\end{equation}
In case of $V=V_{b}$ we receive the rocket's burn time by numerical
integration of
\begin{equation}
t_{b}=\frac{1}{A}\sqrt{\frac{\rho}{2}}\int_{V_{0}}^{V_{b}}\frac{dV}{\sqrt{p_{0}V_{0}^{n}V^{-n}-p_{a}}}.\label{tb-Brennzeit}
\end{equation}
In fact, there is still left some compressed air after the
whole water is expelled. This provides an additional air thrust. Strictly
speaking,  \eqref{tb-Brennzeit} gives the duration of the
water-thrust phase. However, this time will be 
very short. We will discuss the air boost in further detail later
in Section \ref{sec:Movement-in-thrust}.

\subsection{Polytropic index}

\begin{figure}[htb]\centering
\includegraphics[width=0.5\textwidth]{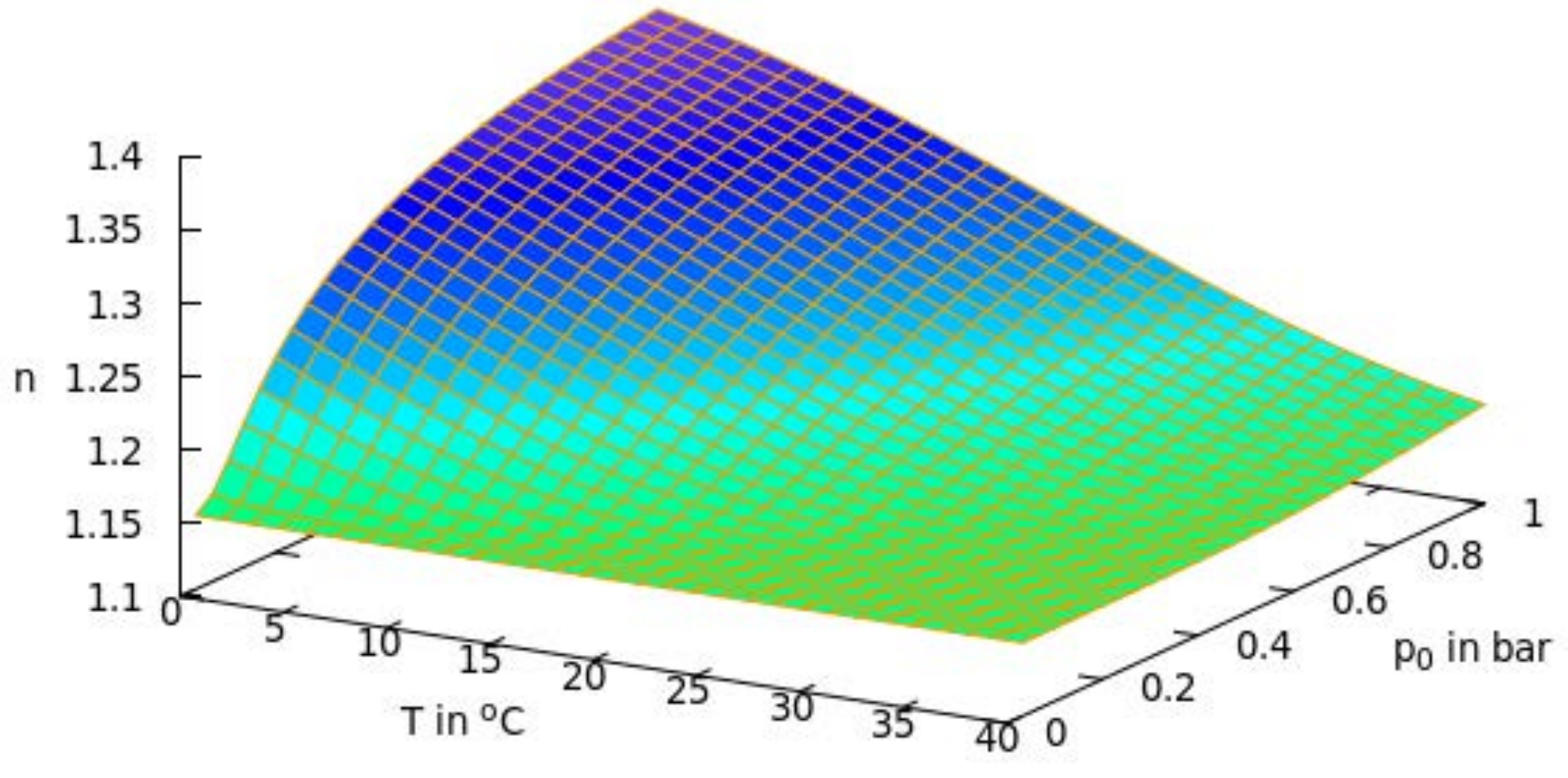}%
\includegraphics[width=0.5\textwidth]{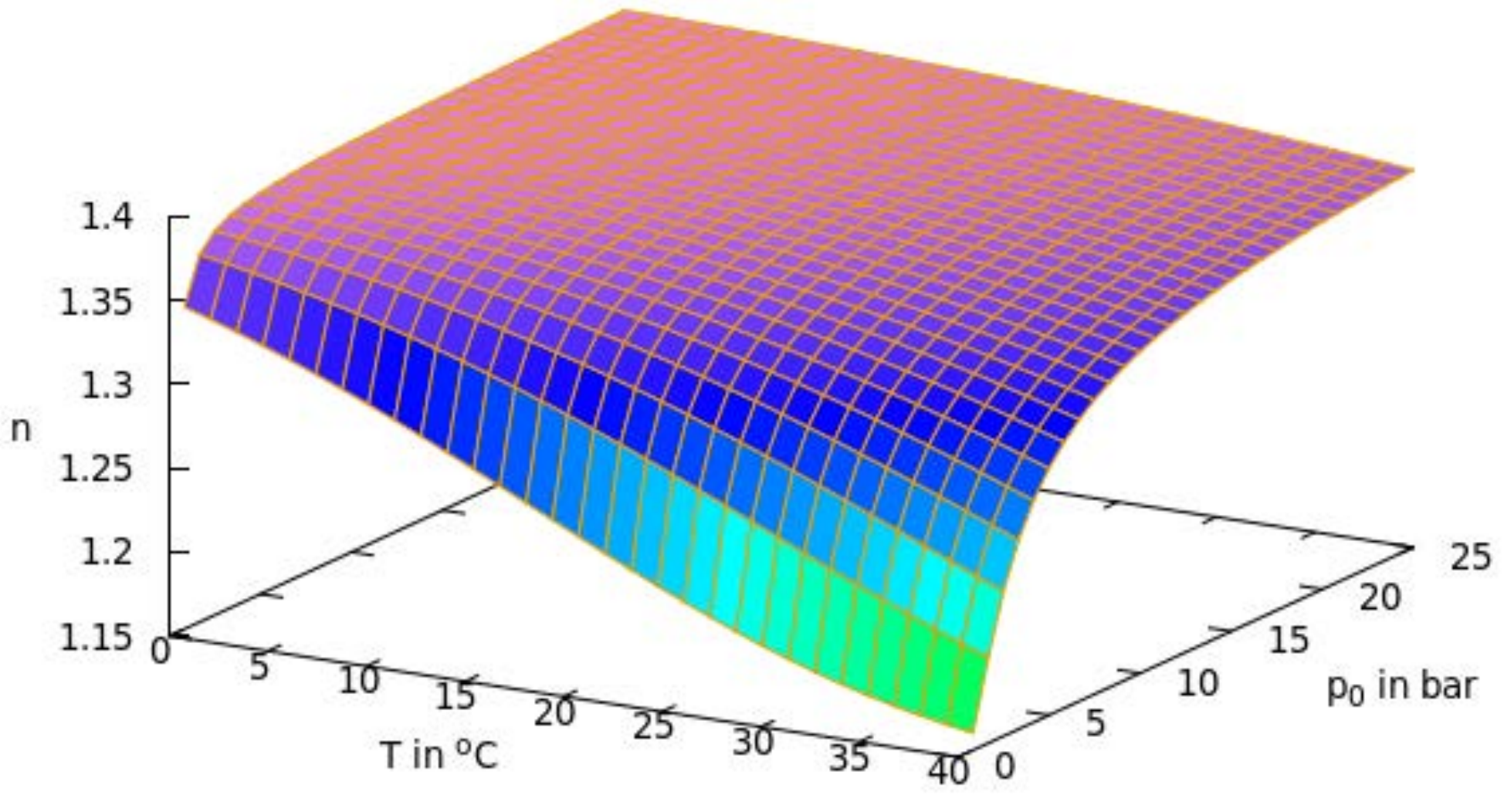}
\caption{\label{fig:Polytropic-index}Polytropic index $n\left(T,p_{0}\right)$
where temperature varies from $0^{\circ}{C}$ to $40^{\circ}{C}$.
The pressure range is $0\leq p_{0}\leq1\,{bar}$ and $0\leq p_{0}\leq25\,{bar}$, respectively.}
\end{figure}

Specific investigation of the gas expansion requires the value of
the polytropic exponent. Frequently, the polytropic index is chosen
as $n=1.4$, see \cite{soda-bottle,Prusa2000} for example. This corresponds
to the assumption that the gas in the rocket is dry air. As mentioned
in \cite{Gommes2010} ``\emph{the air expansion in the rocket is
accompanied by water vapor condensation, which provides an extra thrust}''.
The water vapor pressure at which water vapor is in thermodynamic
equilibrium with the water depends on the temperature. For moist air
the relation of saturation vapor pressure and temperature can be well
approximated by an equation given by Arden Buck in \cite{Buck}. If
$T$ denotes the air temperature in $^{\circ}C$, the saturated vapor
pressure $p_{v}$ is given by 
\begin{equation}\begin{aligned}
&p_{v}=K\cdot\exp\left[\left(18.678-\frac{T}{234.5}\right)\left(\frac{T}{257.14+T}\right)\right]\,,\\ &K=611.21\,\frac{{N}}{m^{2}}=6.1121\cdot10^{-3}{bar}\,.
\end{aligned}\label{Buck}
\end{equation}
In \cite{Air-Expansion-2013,Gommes2010} an approximation of the polytropic
index that depends on the water vapor pressure is given by
\begin{equation}
n=1.15+\left(1.4-1.15\right)\exp\left(-36\frac{p_{v}}{p_{0}}\right).\label{polytrop_n}
\end{equation}
For example, $p_{0}=3\,{bar}$ and $T=15^{\circ}{C}$ yield
$n\approx1.35$. Figure \ref{fig:Polytropic-index} shows the polytropic
index in dependence of temperature and pressure, i.e.\ the function
$n\left(T,p_{0}\right)$ given by \eqref{polytrop_n} together with
\eqref{Buck}. For the expansion of moist air in water rockets the
polytropic index is about $1.1\leq n\leq1.4$.

\subsection{\label{subsec:Analytic-solution-of}Analytic solution of the gas
expansion equation}

\begin{figure}[htb]\centering
\includegraphics[width=0.50\textwidth]{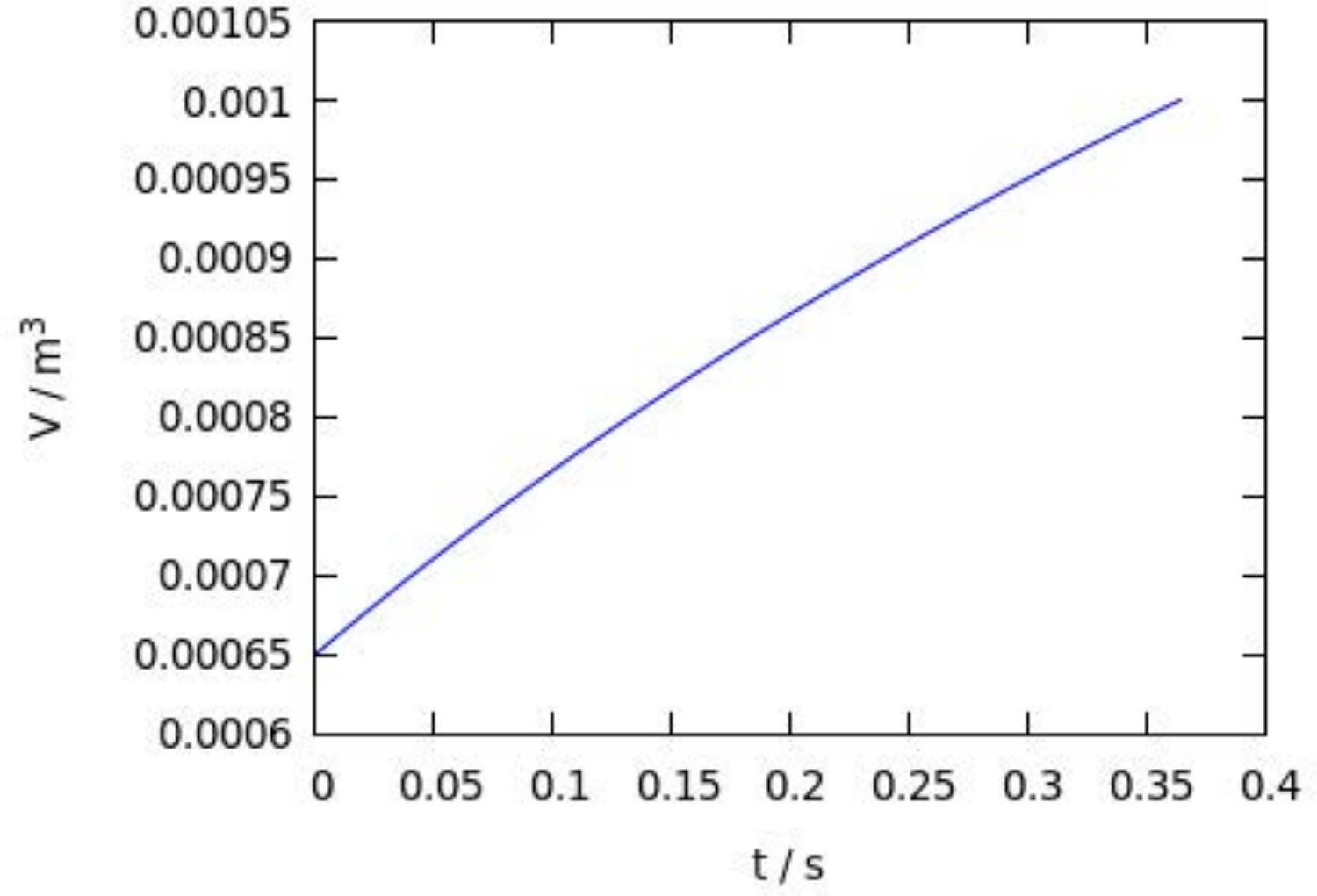}%
\includegraphics[width=0.50\textwidth]{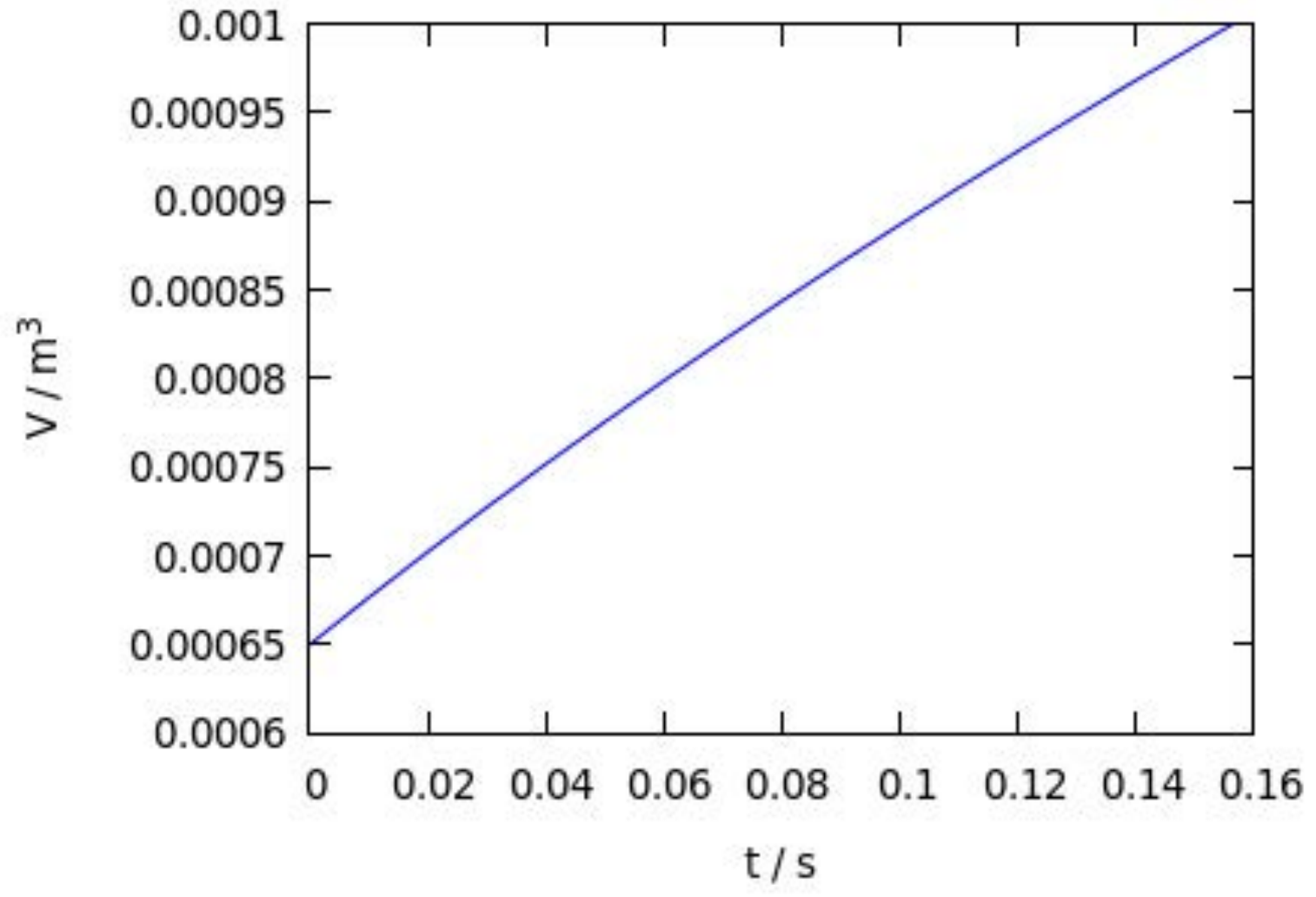}
\caption{\label{fig:Expansion-of-watervapor}Expansion of water vapor from
$0.65\,dm^{3}$ to $1\,dm^{3}$ at $15^{\circ}{C}$.
The left side shows the expansion with an initial pressure of $3\,{bar}$.
This leads to a polytropic exponent of $n\approx1.35$. In the other
case, the initial pressure is $10\,{bar}$, which leads to $n\approx1.39$.
The polytropic exponent was determined by \eqref{polytrop_n}.}
\end{figure}

First, the question arises if there is an analytical solution for
\eqref{dVdt-mit-Polytrop-n}. If that is not the case, the corresponding
equation can be solved numerically. The following approach is based
on the Gaussian hypergeometric function. A little manipulation of
the ordinary differential  \eqref{dVdt-mit-Polytrop-n} for
the volume yields 
\[
V^{\frac{n}{2}}\frac{dV}{dt}=A\sqrt{\frac{2p_{0}V_{0}^{n}}{\rho}}\sqrt{1-\frac{p_{a}}{p_{0}V_{0}^{n}}V^{n}}.
\]
Together with $\frac{d}{dt}\left[V^{\frac{n+2}{2}}\right]=\frac{n+2}{2}V^{\frac{n}{2}}\frac{dV}{dt}$
the latter equation leads to 
\begin{multline*}
\frac{d}{dt}\left[\left(\frac{p_{a}}{p_{0}V_{0}^{n}}\right)^{\frac{n+2}{2n}}V^{\frac{n+2}{2}}\right]\\
=\frac{(n+2)A}{2}\left(\frac{p_{a}}{p_{0}V_{0}^{n}}\right)^{\frac{n+2}{2n}}\sqrt{\frac{2p_{0}V_{0}^{n}}{\rho}}\sqrt{1-\left[\left(\frac{p_{a}}{p_{0}V_{0}^{n}}\right)^{\frac{n+2}{2n}}V^{\frac{n+2}{2}}\right]^{\frac{2n}{n+2}}}\,.
\end{multline*}
By using
\[
u:=\Big(\frac{p_{a}}{p_{0}V_{0}^{n}}\Big)^{\frac{n+2}{2n}}V^{\frac{n+2}{2}}\,,\  k:=\frac{(n+2)A}{2}\Big(\frac{p_{a}}{p_{0}V_{0}^{n}}\Big)^{\frac{n+2}{2n}}\sqrt{\frac{2p_{0}V_{0}^{n}}{\rho}}\,,\ \alpha:=\frac{2n}{n+2}
\]
we finally get 
\begin{equation}
\frac{du}{\sqrt{1-u^{\alpha}}}=k\,dt\,.\label{eq:1-wurzel}
\end{equation}
This equation admits the solution 
\begin{equation}
u\cdot H\big(\tfrac{1}{2},\tfrac{1}{\alpha};\tfrac{1}{\alpha}+1;u^{\alpha}\big)=kt+\xi_{0}\label{eq:2-uH}
\end{equation}
where $\xi_{0}=u_{0}\cdot H\big(\tfrac{1}{2},\tfrac{1}{\alpha};\tfrac{1}{\alpha}+1;u_{0}^{\alpha}\big)$
and $u_{0}=u(0)$. Here $H(a,b;c;x)=\sum\limits _{i=0}^{\infty}\frac{\Gamma(a+i)\Gamma(b+i)\Gamma(c)}{\Gamma(a)\Gamma(b)\Gamma(c+i)i!}x^{i}$
is a special generalized hypergeometric function aka Gauss hypergeometric
function. More precisely, this function is a special solution of Euler's
hypergeometric differential equation, see \cite[eq.\ 1.498]{Bronstein},
\[
x(x-1)w''+\big((a+b+1)x-c)w'+ab\,w=0\,
\]
which can also be written as $\big(x\tfrac{d}{dx}+a\big)\big(x\tfrac{d}{dx}+b\big)w=\big(x\tfrac{d}{dx}+c\big)w'$.
The function $H(a,b;c;x)$ obeys
\begin{align}
\frac{d}{dx}\big(x^{c-1}H(a,b;c;x)\big) & =(c-1)x^{c-2}H(a,b;c-1;x)\,,\label{eq:dHdx}\\
H(a,b;b;x) & =(1-x)^{-a}\,,\label{eq:H-1-x-a}
\end{align}
see \cite[eq.\  1.4.1.6]{Slater1966} and \cite[eq.\ 1.5.1]{Slater1966}.
Therefore, by writing $g(x)=x^{\frac{1}{\alpha}}H\big(\frac{1}{2},\frac{1}{\alpha};\frac{1}{\alpha}+1;x\big)$
and 
 $\tilde{g}(u)=g(x(u))=uH\big(\frac{1}{2},\frac{1}{\alpha};\frac{1}{\alpha}+1;u^{\alpha}\big)$
with $x(u)=u^{\alpha}$ we get 
\begin{equation*}
\begin{aligned}
\frac{d\tilde{g}}{du}&=\frac{dg}{dx}\Big|_{x=u^{\alpha}}\cdot\frac{dx}{du}\;\overset{\eqref{eq:dHdx}}{=}\;\frac{1}{\alpha}x^{\frac{1}{\alpha}-1}H\big(\tfrac{1}{2},\tfrac{1}{\alpha};\tfrac{1}{\alpha};x\big)\Big|_{x=u^{\alpha}}\cdot\alpha u^{\alpha-1}\\&=H\big(\tfrac{1}{2},\tfrac{1}{\alpha};\tfrac{1}{\alpha};u^{\alpha}\big)\;\overset{\eqref{eq:H-1-x-a}}{=}\;\frac{1}{\sqrt{1-u^{\alpha}}}
\end{aligned}
\end{equation*}
as stated. Although a solution of \eqref{dVdt-mit-Polytrop-n} is
represented by  \eqref{eq:2-uH}, its implicit form is inexpedient
for our concerns. Of course, the burn time $t_{b}$ can be calculated
from \eqref{eq:2-uH}. But we also need a closed-form expression for
the gas volume as a function of time. Hence, the expansion equation
\eqref{dVdt-mit-Polytrop-n} will be included into the system of equations
of motion for our rocket. However, let us take a look at the separate
numerical solution of \eqref{dVdt-mit-Polytrop-n} now. Within our
calculations with the Runge Kutta method the temperature of water
and vapor is chosen to be $15^{\circ}{C}$. The volume of the
bottle and the initial water volume are $1\,dm^{3}$ and $0.35\,dm^{3}$,
respectively. The corresponding Figure \ref{fig:Expansion-of-watervapor}
was created with {\tt wxMaxima}. The source code is available at \href{https://github.com/tguent/code}{https://github.com/tguent/code}.

\section{\label{sec:Movement-in-thrust}Movement during water thrust phase}

The flight of the rocket can be divided into three parts: Thrust phase,
coasting phase with upwards motion, and free-fall phase. We model the
thrust phase by dividing it again into two parts: The main thrust
is provided by the expelled water. During the water thrust phase the
acceleration of the water rocket is modeled by \eqref{a_rakete-gesamt}.
But usually there is still some compressed air left at the end of
the water thrust phase. This gives rise to an additional nonzero air
thrust. We refer to this as the air thrust phase in the
following. After all propellant (water and air) is exhausted,
the rocket can be regarded as an object that is thrown vertically
upwards. The corresponding initial velocity is determined by the velocity
at the end of the thrust phase. After reaching the maximum altitude,
the rocket enters the free-fall phase. During the upward coasting and
free-fall phase air drag has a crucial influence on the rocket's movement.
As we will see in the following, air drag is negligible during the
thrust phase of a water rocket. 

\begin{rem}[\bf Example data]\label{Data-of-reference}
For our numerical calculations we use the following data: The empty
rocket has a mass of $1/8\,{kg}$ and a volume of $1\,dm^{3}$,
its nozzle has a diameter of $9\,{mm}$. The rocket radius is
$4\,{cm}$. Initial values for propellant volume and pressure
are $V_{0}=0.35\,dm^{3}$ and $p_{0}=3\,{bar}$ at a temperature
of $15^{\circ}{C}$. The mass of the compressed air is approximately
$1.8\,{g}$. The air drag coefficient was set to $c_{D}=1$. 
\end{rem}

\subsection{\label{subsec:Thrust-phase-including}Water thrust phase including
air drag}

\begin{figure}[htb]\centering
\includegraphics[width=0.50\textwidth]{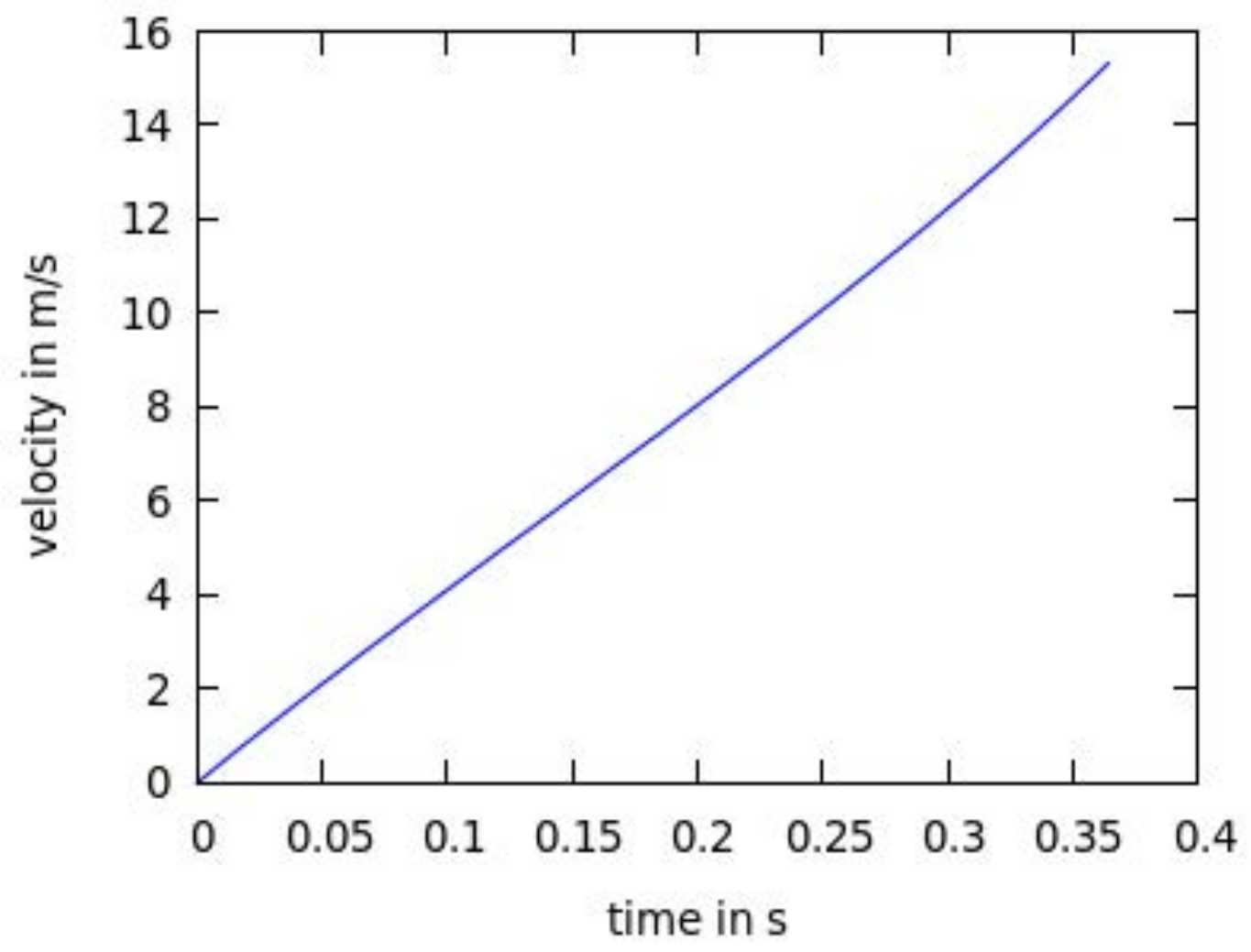}%
\includegraphics[width=0.50\textwidth]{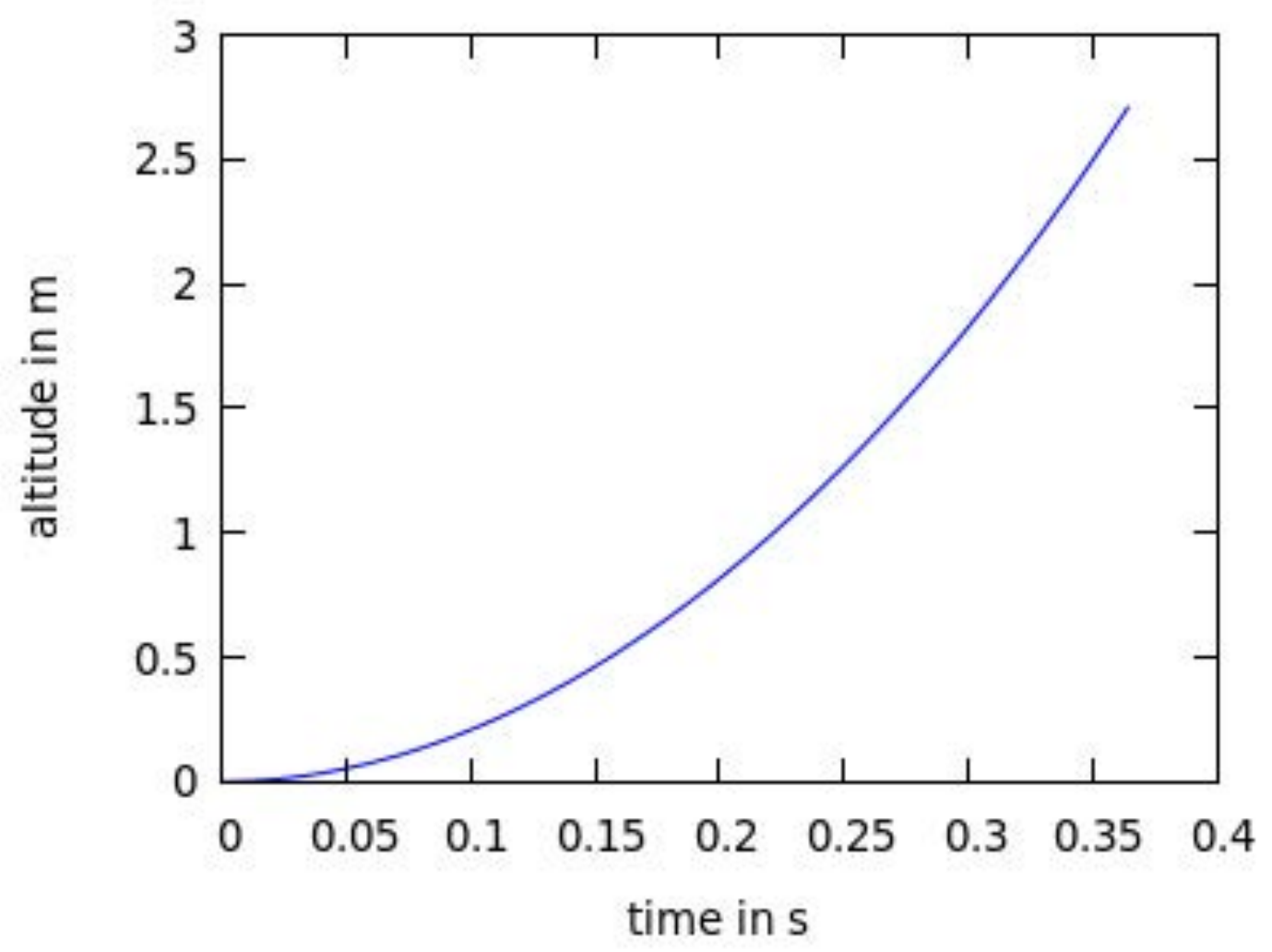}
\caption{\label{fig:Velocity-and-altitude}Velocity and altitude of the rocket
during the water thrust phase for the reference data, see Remark \ref{Data-of-reference}.
At the end of the water thrust phase the rocket has a velocity of
$v_{b}\approx15.32\,\frac{{m}}{{s}}$ at an altitude of
$h_{b}\approx2.71\,{m}$. }
\end{figure}

As mentioned in Section \ref{subsec:Polytropic-expansion-during},
pressure $p$ and volume $V$ are related by polytropic expansion
\eqref{p-polytrop} where $V$ as a function of time $t$ is determined
by  \eqref{dVdt-mit-Polytrop-n}. The rocket's velocity
can be expressed as the rate of change of its position $h$ by $v=\frac{dh}{dt}$.
Analogously, the acceleration \eqref{a-Rakete-Vt} can be expressed
as the rate of change of its velocity $a=\frac{dv}{dt}$. Finally,
this leads to the following first order system of differential equations
for gas volume $V$, altitude $h$ and velocity $v$:
\begin{equation}
\begin{gathered}
\frac{dV}{dt}=A\sqrt{\frac{2\left(p_{0}V_{0}^{n}V^{-n}-p_{a}\right)}{\rho}}\,,\quad 
\frac{dh}{dt}=v\,, \\
\frac{dv}{dt}=\frac{2A\left(p_{0}V_{0}^{n}V^{-n}-p_{a}\right)-Dv^{2}}{M+m_{\text{air}}+\rho V_{b}-\rho V}-g\,.
\end{gathered}
\label{DGL-SYSTEM-V-v-h}
\end{equation}
The latter can be solved numerically. We use the fourth order Runge
Kutta method, see \cite{Abramowitz}, which is implemented in the
computer algebra system {\tt wxMaxima}, see \cite{MAXIMA}. The step-size was
set to $0.01$. The corresponding numerical results for the reference
data given in Remark \ref{Data-of-reference} are presented in Figure
\ref{fig:Velocity-and-altitude}. In order to create comparability
we use the following reference values $v_{b}\approx15.32\,\frac{{m}}{{s}}$
at an altitude of $h_{b}\approx2.71\,{m}$ at the end of the
water thrust phase. For the corresponding {\tt wxMaxima} source code see
again \href{https://github.com/tguent/code}{https://github.com/tguent/code}.

\subsection{\label{subsec:romb-approximation}Simple estimate - Water thrust
phase approximation with zero drag coefficient}

Solving the equations of motion \eqref{DGL-SYSTEM-V-v-h} require
advanced numerical calculations. In the following we present a more
simple estimation of altitude and velocity at the end of the water
thrust phase. The first equation of \eqref{DGL-SYSTEM-V-v-h} determines
the gas expansion. Figure \ref{fig:Expansion-of-watervapor} shows
the gas expansion for $0\leq t\leq t_{b}$, where $t_{b}$ is determined
by \eqref{tb-Brennzeit}. This indicates that the gas volume as a
function of time is not too far from being linear, at least as a rough
estimate. If we use 
\begin{equation}
V\left(t\right)=\frac{V_{b}-V_{0}}{t_{b}}\,t+V_{0}\label{Volume-t-linear}
\end{equation}
for our calculations the corresponding equation\footnote{With \eqref{Volume-t-linear} the equations of motion \eqref{DGL-SYSTEM-V-v-h}
reduce to $\frac{dh}{dt}=v\,,\;\frac{dv}{dt}=\frac{2A\left[\frac{p_{0}V_{0}^{n}}{\left(\frac{V_{b}-V_{0}}{t_{b}}\,t+V_{0}\right)^{n}}-p_{a}\right]-Dv^{2}}{M+m_{\text{air}}+\rho V_{b}-\rho\left(\frac{V_{b}-V_{0}}{t_{b}}\,t+V_{0}\right)}-g$.} of motion predicts a velocity of about $15.17\,{{\frac{m}{s}}}$
at an altitude of $2.70\,{m}$ at the end of the thrust phase
(again for the data given in Remark \ref{Data-of-reference}). This
is only a minor deviation downwards from the more accurate reference
values of $v_{b}\approx15.32\,\frac{{m}}{{s}}$ and $h_{b}\approx2.71\,{m}$.
The relative deviation of velocity and altitude at the end of the
thrust phase is about 0.9\% and 0.2\%, respectively.
Thus, the linear approach \eqref{Volume-t-linear} can be used as a
good approximation. Now consider the last equation of \eqref{DGL-SYSTEM-V-v-h}
which incorporates thrust and air resistance. Although air drag has
crucial influence on the maximum altitude of the rocket it has few
influence during the thrust phase. During the subsequent free flight
phase when the propellant is exhausted it is very important to incorporate
air drag again. Next we calculate the data at the end of the water
thrust phase in case of $c_{D}=0$. Together with \eqref{Volume-t-linear},
equation \eqref{DGL-SYSTEM-V-v-h} leads to
\begin{equation}
v_{n}\left(t\right)=\int_{0}^{t}\frac{2A\left[\frac{p_{0}V_{0}^{n}}{\left(\frac{V_{b}-V_{0}}{t_{b}}\,x+V_{0}\right)^{n}}-p_{a}\right]}{M+m_{\text{air}}+\rho V_{b}-\rho\left(\frac{V_{b}-V_{0}}{t_{b}}\,x+V_{0}\right)}dx-gt\label{Estimation-velocity}
\end{equation}
at a given time $t\leq t_{b}$. Integration over the burn time $t_{b}$,
see \eqref{tb-Brennzeit}, gives the velocity $v_{nb}$ at the end
of the water thrust phase. The altitude at the end of a water rocket's
thrust phase is small. We use the mean acceleration $\bar{a}=v_{nb}/t_{b}$
and 
\begin{equation}
h_{nb}:=\frac{1}{2}\bar{a}t_{b}^{2}=\frac{v_{nb}t_{b}}{2}\label{Estimation-altitude}
\end{equation}
as a rough estimate of the altitude at the end of the water
thrust phase. 
With the above used data from Remark \ref{Data-of-reference},
this leads to $v_{nb}\approx15.60\,\frac{{m}}{{s}}$
for $h_{nb}\approx2.84\,{m}$. The results exceed the reference values $v_{b}\approx15.32\,\frac{{m}}{{s}}$
and $h_{b}\approx2.71\,{m}$ by $1.8\%$ and $4,9\%$, respectively.
Such a simple analysis of the motion of a water rocket can actually
be carried out in undergraduate physics courses. All necessary calculations
for the method figured out in Section \ref{subsec:romb-approximation}
can be done with a graphing calculator. 

\section{\label{sec:The-air-thrust}Movement during air thrust phase}

\subsection{Preliminarily remarks}

As mentioned before, the mass of air inside the rocket is
small, i.e. $m_{\text{air}}\ll M$. Even if we neglect the mass of
the compressed air during the water thrust phase the error will be
very small. Using \eqref{p-polytrop} the mass of air inside the rocket
can be calculated from
\begin{equation}
m_{\text{air}}=\rho_{\text{air}}\sqrt[n]{\frac{p_{0}}{p_{a}}}\:V_{0}\label{m_air}
\end{equation}
where $\rho_{\text{air}}\approx1.23\frac{g}{dm^{3}}$ is the
density of air at atmospheric pressure $p_{a}\approx1\,{bar}$.
For example: An initial air volume of $V_{0}=0.65\,dm^{3}$
at $p_{0}=3\,{bar}$ contributes to the total mass of the rocket
with about $2\,{g}$. In our case, the rocket has a curb weight
of about $1/8\,{kg}$ plus $350\,{g}$ propellant (water).
The additional mass of $2g$ leads to a deviation of about some centimeters of calculated altitude. 
I.e., the additional mass
of the compressed air doesn't has a noticeable influence on the water
thrust phase where the rocket is propelled by the water. But what
effect has the expelled air after the water thrust phase? On the one
hand, there is only a tiny working mass left to induce a change in
momentum. On the other hand, the air escapes very rapidly. Indeed,
it already was a matter of discussion if the air boost can be neglected
or not. Prusa notes: ``\emph{Due to its low density, the thrust provided
by air alone is negligible, and in a launch without water, the rocket
is barely able to lift off of the air pump seal.}'', cf.$\negthinspace$
\cite[p.\ 719]{Prusa2000}. On the other hand, Barrio-Perotti et.$\negthinspace$
al.$\negthinspace$ state: ``\emph{This air is expelled through the
nozzle causing an additional increase in the rocket momentum that
sometimes cannot be neglected: for example, a rocket with air pressurized
at 2 bars can reach a distance of about 10 m }{[}...{]}'', cf.$\negthinspace$
\cite[p.\ 1138]{Barrio2010}. In \cite{Barrio2010}, the air propulsion
is also described by a system of differential equations, see \cite[eq.\ 29, 30, 31]{Barrio2010}
together with the algebraic equation \cite[eq.\ 28]{Barrio2010} which
have to be solved numerically. In order to decide whether the air
boost has to be taken into account or not we filled our model rocket
with compressed air (no water) at about $2\,{bar}$ pressure.
Actually, the rocket lifted off and reached a significant altitude
of at least two meters. Thus we decided not to neglect the air propulsion.
But since the effect is small compared to the water thrust, we choose
a straightforward approximation of the air thrust phase in the following. 

\subsection{Rocket equation with simplified model assumptions}

The gas volume equals the bottle's volume $V_{b}$ now. According 
to \eqref{p-polytrop}, initial pressure $p_{0}$ and initial air
density have reduced to
\begin{equation}
p_{b}=p_{0}\left(\frac{V_{0}}{V_{b}}\right)^{n}\quad\text{and}\quad\rho_{b}=\frac{m_{\text{air}}}{V_{b}}\;\overset{\eqref{m_air}}{=}\;\rho_{\text{air}}\frac{V_{0}}{V_{b}}\,\sqrt[n]{\frac{p_{0}}{p_{a}}}\label{pb-und-rhob}
\end{equation}
The change in momentum can be calculated from the rocket equation
\eqref{a_rakete-gesamt}. In order to avoid complicated numerical
methods we neglect the air drag during the air thrust phase. Barrio-Perotti
et.\ al.\ evaluate water and air thrust in \cite{Barrio2010}. In comparison
with the duration of the water thrust phase, the air thrust phase
is short. Beyond that, the rocket acceleration is strongly decreasing
during the air propulsion, see \cite[Fig.\ 12]{Barrio2010}: If the
air propulsion lasts about $0.025$ seconds, the major contribution
may be over within a hundredth of a second. Compared with this, the
water thrust in \cite[Fig.\ 12]{Barrio2010} lasts six times longer
(and leads to higher acceleration anyway). Based on the short duration
of the air propulsion, we assume that the exhaust velocity $v_{e}$
can be treated as a constant. With these simplifications the rocket
equation $\frac{dv}{dt}=-\frac{v_{e}dm}{m\,dt}-g$ can be integrated.
\begin{equation}
\int_{v_{b}}^{v_{\text{max}}}dv=-v_{e}\int_{M+m_{\text{air}}}^{M}\frac{dm}{m}-\int_{\text{duration of air propulsion}}g\,dt.\label{eq:intergal-eq-air-thurst}
\end{equation}
Now let us rate the two terms on the right side of \eqref{eq:intergal-eq-air-thurst}.
Because the change of mass can be huge the integral over the mass
may have a significant contribution to the result, despite of the
fact that the duration of the air propulsion is very short. Now consider
the last integral. The product of gravitational acceleration and duration
of air propulsion will be also small. So we neglect this part. The
remaining terms result in $v_{\text{max}}\approx v_{e}\ln\left(\frac{M+m_{\text{air}}}{M}\right)+v_{b}$
where $v_{b}$ is the rocket velocity at the end of the water thrust
phase. Indeed, the above considerations are further based on the assumption
that no air escapes before all water is expelled. This seems
to be unrealistic. Particularly towards the end of the water thrust
phase the propellant will be a mixture of water and air. Therefore,
the mass of the remaining air at the end of water thrust will be less
than the initial amount of air inside the rocket given by \eqref{m_air}.
In order to take into account the loss of air during the water thrust
phase we include an efficiency factor $\eta$ and receive 
\begin{equation}
v_{\text{max}}=v_{e}\ln\left(\frac{M+\eta\cdot m_{\text{air}}}{M}\right)+v_{b}.\label{vmax-mit-eta-und-ve}
\end{equation}
Amongst other things, the efficiency factor $\eta$ presumably depends
on the initial proportion of water and air in the rocket, that is
$V_{0}/V_{b}$. If the initial amount of air equals the bottle volume,
there is no water propulsion and the air provides the whole thrust.
Obviously the air thrust efficiency should be $\eta=1$ in this case.
If the whole bottle is filled with water, $\eta=0$ should indicate
that there will be no air thrust. Furthermore, the efficiency strongly
depends on the construction of the D.I.Y rocket. Since we consider
a highly chaotic system, there should be a parameter to readjust our
calculation to the experimental data of a special model rocket. Based
on the preceding considerations we choose
\begin{equation}
\eta=\left(\frac{V_{0}}{V_{b}}\right)^{\mu}.\label{eta-efficiency-factor}
\end{equation}
As a first estimate one can use $\mu=1$. In
order to fine-tune this estimate for a special D.I.Y. rocket, the
exponent $\mu\geq0$ can be determined from the experimental data:
For our rocket a value of $\mu\approx3$ works quite well.

Now we still need the exhaust velocity $v_{e}$. We cannot simply
calculate $v_{e}$ from \eqref{ve}, since it is based on Bernoulli's
equation for incompressible flow. The density of a compressible flow
is not constant. But the derivation of Bernoulli's equation from $\frac{dp}{\rho}+v\,dv+g\,dz=0$
can be modified for compressible flow. With $p\propto\rho^{n}$, see
\cite[eq.\ 3.20]{Aerodynamics-Clancy}, we receive the relationship
$\frac{n}{n-1}\cdot\frac{p}{\rho}+\frac{1}{2}v^{2}=constant$ which
applies to compressible adiabatic flows, see \cite[eq. 3.21]{Aerodynamics-Clancy}.
We use again the above notation $\rho_{\text{air}}$ for the air density
at atmospheric pressure $p_{a}$. Density and pressure at the beginning
of the air thrust phase are denoted by $\rho_{b}$ and $p_{b}$, respectively.
Let us assume that the speed inside the bottle can be neglected. Therewith,
we receive for the exhaust velocity: 
\begin{equation}
v_{e}=\sqrt{\frac{2n}{n-1}\left(\frac{p_{b}}{\rho_{b}}-\frac{p_{a}}{\rho_{\text{air}}}\right)}\;\overset{\eqref{pb-und-rhob}}{=}\;\sqrt{\frac{2n\,p_{a}}{\left(n-1\right)\rho_{\text{air}}}\left[\left(\frac{p_{0}}{p_{a}}\right)^{1-\frac{1}{n}}\left(\frac{V_{0}}{V_{b}}\right)^{n-1}-1\right]}.\label{ve-air-thrust}
\end{equation}
Combining \eqref{vmax-mit-eta-und-ve} and \eqref{ve-air-thrust}
results in 
\begin{equation}
v_{\text{max}}\approx\sqrt{\frac{2n\,p_{a}}{\left(n-1\right)\rho_{\text{air}}}\left[\left(\frac{p_{0}}{p_{a}}\right)^{1-\frac{1}{n}}\left(\frac{V_{0}}{V_{b}}\right)^{n-1}-1\right]}\,\ln\left(\frac{M+\eta\cdot m_{\text{air}}}{M}\right)+v_{b}.\label{vmax-air-thrust-formel}
\end{equation}

Since the duration of the air propulsion is very short, we neglect
the additional height which the rocket attains during this phase.
But we take into account the enhanced velocity from the air thrust.
This significantly affects the upward coasting. 

\section{\label{sec:Upward-Coast}Movement in the coasting phase}

By the time the working mass is expelled the rocket has already reached
its top speed. From this point we can regard the rocket as an object
that is thrown vertically upwards. Air resistance has a significant
impact on the movement after the thrust phase. Luckily, the corresponding
equations of motion can be solved analytically.

The kinematics of an upward movement including air resistance are
well known. As the rocket's propellant is exhausted the remaining
mass $M$ of the rocket is constant. With the notation
\begin{equation}
\psi=\frac{D}{M}=\frac{\rho_{\text{air}}\,c_{D}\,A_{R}}{2M},\label{psi-def}
\end{equation}
the air drag deceleration \eqref{aD-air-drag}
takes the form $a_{D}=\psi v^{2}$ during the free flight phase, see \eqref{D-def}. For
$v\leq v_{\text{max}}$ the equation of motion $\frac{dv}{dt}=-\left(g+\psi v^{2}\right)$
leads to the integral equation 
\[
\int_{v_{\text{max}}}^{v}\frac{dx}{g+\psi x^{2}}=-\int_{0}^{t}dt
\]
where $t=0$ corresponds to the end of the thrust phase. The solution
is given by 
\begin{equation}
v\left(t\right)=\sqrt{\frac{g}{\psi}}\tan\left(\arctan\left(\sqrt{\frac{\psi}{g}}\,v_{\text{max}}\right)-\sqrt{\psi g}\:t\right).\label{v-luft}
\end{equation}
With $v=0$ the duration of the upward coasting phase is given by 
\begin{equation}
t_{c}=\frac{1}{\sqrt{\psi g}}\:\arctan\left(\sqrt{\frac{\psi}{g}}\,v_{\text{max}}\right).\label{tc}
\end{equation}
The upward coasting begins after the thrust phase and ends when the rocket
has reached the maximum altitude. The covered distance during the
free flight phase is given by 
\begin{align*}
h_{c} & =\int_{0}^{t_{c}}\sqrt{\frac{g}{\psi}}\tan\left(\arctan\left(\sqrt{\frac{\psi}{g}}\,v_{\text{max}}\right)-\sqrt{\psi g}\:t\right)\,dt\\
 & =\frac{1}{\psi}\left\{ \ln\left|\cos\left(\arctan\left(\sqrt{\frac{\psi}{g}}\,v_{\text{max}}\right)-\sqrt{\psi g}\:t_{c}\right)\right|-\ln\left|\cos\left(\arctan\left(\sqrt{\frac{\psi}{g}}\,v_{\text{max}}\right)\right)\right|\right\} \\
 & =-\frac{1}{\psi}\ln\left|\cos\left(\arctan\left(\sqrt{\frac{\psi}{g}}\,v_{\text{max}}\right)\right)\right|.
\end{align*}
The first term in the second row cancels out by using  \eqref{tc}
for $t_{c}$. Using $\cos\left(\arctan\,x\right)=\frac{1}{\sqrt{1+x^{2}}}$,
i.e.\footnote{The absolute value function can be abandoned since
$x>0$.}
\[
\ln\left|\cos\left(\arctan\left(x\right)\right)\right|=\ln\left(\frac{1}{\sqrt{1+x^{2}}}\right)=-\frac{1}{2}\ln\left(1+x^{2}\right),
\]
one finally gets 
\begin{equation}
h_{c}=\frac{1}{2\psi}\ln\left(1+\frac{\psi}{g}\,v_{\text{max}}^{2}\right)\label{eq:hc_coast-phase}
\end{equation}
where $\psi=\frac{\rho_{\text{air}}\,c_{D}\,A_{R}}{2M}$, see \eqref{psi-def}.

\section{\label{sec:Collection-of-theResults}Collection of the results and
maximum altitude}

Let us summarize the previous results and therewith derive a method
to estimate the maximum altitude of the rocket. First we have to calculate
the burn time of the water thrust phase by 
\[
t_{b}=\frac{1}{A}\sqrt{\frac{\rho}{2}}\int_{V_{0}}^{V_{b}}\frac{dV}{\sqrt{p_{0}V_{0}^{n}V^{-n}-p_{a}}},
\]
see  \eqref{tb-Brennzeit}. Here $A$ is the nozzle cross
sectional area, $\rho$ the water density, $n$ the polytropic exponent,
$p_{a}$ the atmospheric pressure and $p_{0}$ the initial pressure
in the rocket tank. The tank has the volume $V_{b}$. At launch, $V_{0}<V_{b}$
is filled with air and $V_{b}-V_{0}$ is the initial water volume.
The water thrust phase can be modeled by the system of differential
equations \eqref{DGL-SYSTEM-V-v-h}. Its numerical solution gives
velocity $v_{b}$ and altitude $h_{b}$ at the end of the water thrust
phase. Alternatively, for a rough estimate we receive these values
from 
\begin{equation}
\begin{aligned}
v_{b} & \approx v_{nb}=\int_{0}^{t_{b}}\frac{2A\left[\frac{p_{0}V_{0}^{n}}{\left(\frac{V_{b}-V_{0}}{t_{b}}\,x+V_{0}\right)^{n}}-p_{a}\right]}{M+m_{\text{air}}+\rho V_{b}-\rho\left(\frac{V_{b}-V_{0}}{t_{b}}\,x+V_{0}\right)}dx-gt_{b}\,,\\  
h_{b}& \approx h_{nb}=\frac{v_{b}t_{b}}{2}\,,
\end{aligned}\label{eq:3}
\end{equation}
see \eqref{Estimation-velocity} and \eqref{Estimation-altitude}.
$M$ is the rocket's curb mass, $m_{\text{air}}$ the initial
mass of air in the tank and $g$ the gravitational acceleration.
After the water thrust phase, the rocket experiences an additional
acceleration due to the air propulsion. We only consider the enhanced
velocity \eqref{vmax-air-thrust-formel} within our model
\[
v_{\text{max}}\approx\sqrt{\frac{2n\,p_{a}}{\left(n-1\right)\rho_{\text{air}}}\left[\left(\frac{p_{0}}{p_{a}}\right)^{1-\frac{1}{n}}\left(\frac{V_{0}}{V_{b}}\right)^{n-1}-1\right]}\,\ln\left(\frac{M+\eta\cdot m_{\text{air}}}{M}\right)+v_{b}
\]
where $\eta$ is the efficiency factor of air propulsion, see
 \eqref{eta-efficiency-factor}. The rocket enters the upward
coasting phase with the velocity $v_{\text{max}}$ at the altitude $h_{b}$.
During the upward coasting the rocket reaches an additional height of
\[
h_{c}=\frac{M}{\rho_{\text{air}}\,c_{D}\,A_{R}}\ln\left(1+\frac{\rho_{\text{air}}\,c_{D}\,A_{R}}{2Mg}\,v_{\text{max}}^{2}\right)\,,
\]
see \eqref{eq:hc_coast-phase}. Finally, the maximum altitude of the
rocket can be calculated from  $h_{\text{max}}=h_{b}+h_{c}$.

\section{\label{sec:Drag-analysis}Drag analysis - An estimate of the drag
coefficient}

The drag coefficient can be divided into its components which arise
from pressure drag and friction drag, see \cite[eq.\ 7.63]{White-Fluid-Mechanics}.
Friction drag is caused by the viscosity of the surrounding air in
our case. Pressure drag is the ``\emph{difference between the high
pressure in the front stagnation region and the low pressure in the
rear separated region} {[}...{]}'', cf.\ \cite[p.\ 448]{White-Fluid-Mechanics}.
Basically, the drag coefficient varies with the Reynolds number $Re=v\,L/\nu$
where $v$ is the rocket's velocity, $L$ is its characteristic
length, and $\nu$ is the kinematic viscosity\footnote{The kinematic viscosity of air at $15^{\circ}C$ is $\approx1,5\cdot10^{-5}\frac{m^{2}}{{s}}$.
The NASA provides a ``Similarity Parameter Calculator'' which calculates
the Reynolds number: \href{https://www.grc.nasa.gov/www/k-12/airplane/viscosity.html}{https://www.grc.nasa.gov/www/k-12/airplane/viscosity.html} } of the surrounding atmosphere, see \cite[eq.\ 7.61]{White-Fluid-Mechanics}.
A simple home-made rocket built from a PET bottle achieves a maximum
speed of about $20\frac{{m}}{{s}}$. 
From $Re=v\,L/\nu$ we see that the Reynolds number will not exceed
$5\cdot10^{5}$ in this case. Of course, a more professional constructed
water rocket might receive a maximum Reynolds number which is about
ten times higher. The relation\footnote{It is worth mentioning that the drag force $F_{D}=\frac{1}{2}\rho_{\text{air}}c_{D}Av^{2}$
increases with increasing speed of the rocket. Although if the drag
coefficient $c_{D}$ usually decreases. } of drag coefficient and Reynolds number is usually obtained from
laboratory experiments, see for example \cite[fig.\ 7.16]{White-Fluid-Mechanics}.
Since such precise aerodynamic considerations are beyond the scope
of this paper we try to get an estimate of a constant drag coefficient of our
rocket in the following. ``\emph{The drag analysis of rockets} {[}...{]}
\emph{is usually simplified by considering the rocket to be made up
of several simple basic components}'' \cite{Aerodynamic-Drag-Model-Rockets}.
We will confine our considerations to the drag analysis of nose cone,
body tube, base, fins, and a (small) constant value for the launch
lugs. The latter segmentation of a model rocket is shown in \cite[fig.\ 17]{Aerodynamic-Drag-Model-Rockets}.
Due to interference drag the total drag of the rocket amounts to more
than the sum of the components: ``{[}...{]} \emph{additional amount
of drag is caused by the joining of the fins to the rocket body.}
{[}...{]} \emph{Interference drag can be as much as $10\%$ above
the sum of the fin and body tube drag}'', cf.\ \cite[p.\ 9]{Aerodynamic-Drag-Model-Rockets}.
In this paper, the drag coefficients of nose cone, body tube, base,
fins, interference and launch lugs are denoted by $c_{\text{nose}}$,
$c_{\text{tube}}$, $c_{\text{base}}$, $c_{\text{fin}}$, $c_{\text{int}}$,
and $c_{\text{lau}}$ respectively. The total drag coefficient $c_{D}=c_{\text{nose}}+c_{\text{tube}}+c_{\text{base}}+c_{\text{fin}}+c_{\text{int}}+c_{\text{lau}}$
represents the case that the rocket is moving at zero angle to the
wind direction. Any nonzero angle leads to an additional induced
drag. 

\subsection{Nose cone and body tube of the rocket}

The nose cone is exposed to pressure drag and skin friction drag. As
mentioned in \cite[p.\ 10]{Aerodynamic-Drag-Model-Rockets} a flat
nose cone (solely) would result in a drag coefficient of $c_{N}=0.8$
due to pressure. Gregorek compares the latter case to the
order of magnitude of drag coefficients of several shapes. 
\cite[fig.~23]{Aerodynamic-Drag-Model-Rockets} indicates that rounding
the nose reduces the corresponding drag coefficient by about $90\%$
or even more. Nose cone and body tube of the rocket are additionally
exposed to friction drag. Luckily, there exists an equation, see \cite[eq.\ 8]{Aerodynamic-Drag-Model-Rockets},
which includes the drag of the nose cone as well as the body tube.
Let $A_{cs}$ be the cross-sectional area of the body tube and $A_{ws}$
the wetted surface area of the rocket. Therewith, \cite[eq.\ 8]{Aerodynamic-Drag-Model-Rockets}
takes the form
\begin{equation}
\mathrm{\mathcal{C}_{l}}:=c_{\text{nose}}+c_{\text{tube}}=1.02\,c_{f}\left(1+\frac{3}{2\left(\frac{L}{d}\right)^{\frac{3}{2}}}\right)\frac{A_{ws}}{A_{cs}}\label{c_nose-tube}
\end{equation}
where $c_{f}$ is the skin friction coefficient and $L/d$ the length
to diameter ratio of the rocket. 

\subsection{Base and fins of the rocket}

\begin{figure}[htb]\centering
\includegraphics[width=0.50\textwidth]{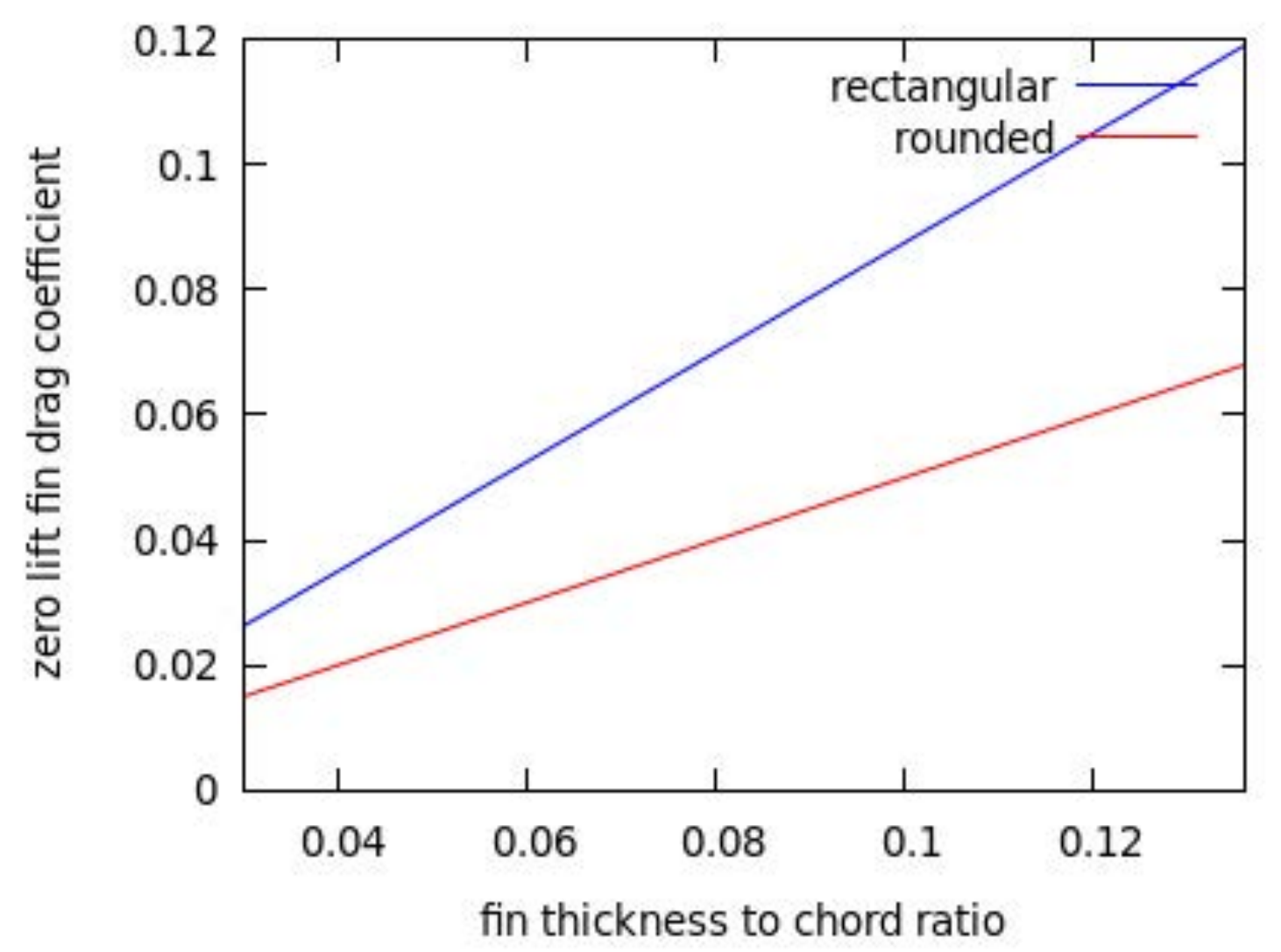}%
\includegraphics[width=0.50\textwidth]{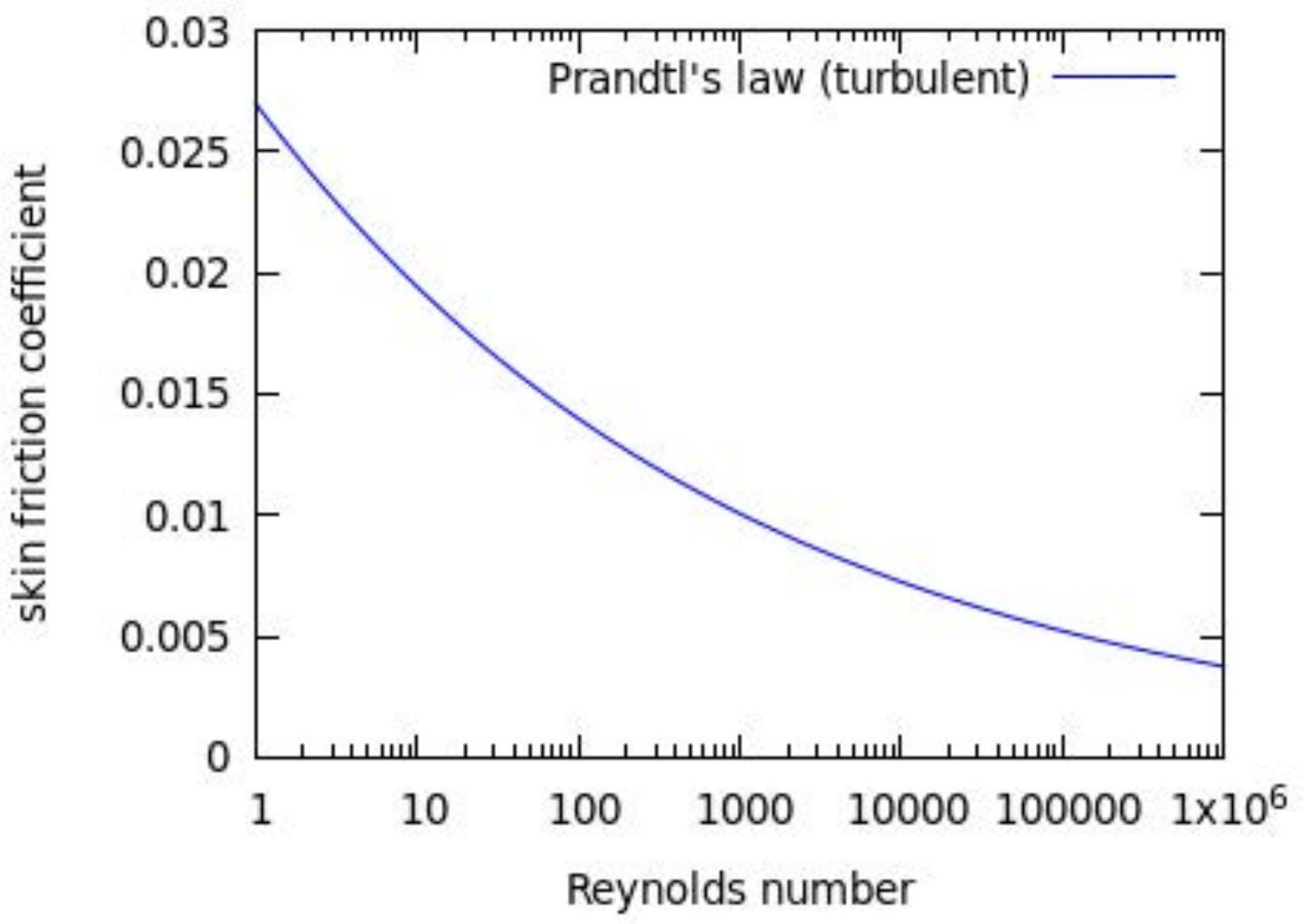}
\caption{\label{fig:Zero-lift-fin-and-skin-friction}Left side: Zero lift fin
drag coefficients for fins with rectangular and rounded cross-sections.
Right side: Skin friction coefficient as a function of Reynolds number
given by Prandtl's law, see \cite[eq.\ 7.43]{White-Fluid-Mechanics}. }
\end{figure}

Flow separation causes low pressure at the rear of the rocket which
results in base drag. An equation that estimates the base drag is
given by 
\begin{equation}
c_{\text{base}}=\frac{0.029}{\sqrt{c_{\text{nose}}+c_{\text{tube}}}},\label{c_base}
\end{equation}
see \cite[eq.\ 9]{Aerodynamic-Drag-Model-Rockets}. Fins cause additional
friction drag, pressure drag and induced drag. But they substantially
enhance the flight stability of our rocket. Generally, the fin drag
depends on various parameters like the thickness to chord ratio, planform
area and cross-sections. We present a rough estimate in this
paper. In order to get some upper limit we use fairly pessimistic
assumptions for the zero lift fin drag coefficient. Following \cite[p.\ 17]{Aerodynamic-Drag-Model-Rockets},
the fin thickness to chord ratio will rarely be greater than $0.1$
for a typical model rocket. Average values for the zero lift fin drag
coefficient are given in \cite[fig.\ 40]{Aerodynamic-Drag-Model-Rockets}
for rectangular, rounded, and steamlined cross-sections. Certainly,
we cannot choose a steamlined\footnote{According to \cite[fig.\ 41]{Aerodynamic-Drag-Model-Rockets}, the
zero lift fin drag coefficient for steamlined cross-sections at ratio
$0.1$ doesn't exceed $0.019$ (laminar) to $0.024$ (turbulent)
for $30\frac{{m}}{{s}}$ ($100\frac{{ft}}{{sec}}$). } cross-section for a pessimistic estimate. Let $\tau$ denote the
thickness to chord ratio and $c_{\text{fin}}^{*}$ the zero lift fin
drag coefficient in the following. The plotted data in \cite[fig.\ 40]{Aerodynamic-Drag-Model-Rockets}
suggests that $c_{\text{fin}}^{*}$ depends linearly on $\tau$ in
for $0.03<\tau<0.136$. Within this interval we deduced the
following relations:
\begin{align}
c_{\text{fin}}^{*} & \approx0.875\tau\qquad\quad\text{(rectangular cross-section)},\label{cfin-stern-rec}\\
c_{\text{fin}}^{*} & \approx0.5\tau\qquad\quad\quad\text{(rounded cross-section)}\label{cfin-stern-rounded}
\end{align}
The left side of figure \ref{fig:Zero-lift-fin-and-skin-friction}
shows that the graphic illustration of \eqref{cfin-stern-rec} and
\eqref{cfin-stern-rounded} reproduces the corresponding lines in
\cite[fig.\ 40]{Aerodynamic-Drag-Model-Rockets}. 
Let us assume in the following that the fins have rectangular cross-section.
The zero lift fin drag coefficient $c_{\text{fin}}^{*}$ is based
on the fin area $A_{\text{fin}}$ whereas the other drag coefficients
are based on the body tube cross-sectional area $A_{cs}$. Therefore,
we have to adjust the coefficient $c_{\text{fin}}^{*}$ by 
\begin{equation}
c_{\text{fin}}=c_{\text{fin}}^{*}\frac{A_{\text{fin}}}{A_{cs}}\;\overset{\eqref{cfin-stern-rec}}{=}\;0.875\,\frac{A_{\text{fin}}}{A_{cs}}\tau.\label{c_fin}
\end{equation}
As above mentioned, rocket body and fins jointly cause additional
intereferenz drag. Due to \cite[eq.\ 21]{Aerodynamic-Drag-Model-Rockets},
the interference drag can be estimated by 
\begin{equation}
c_{\text{int}}=c_{\text{fin}}^{*}\frac{C_{R}\,d}{2A_{cs}}\cdot N\;\overset{\eqref{cfin-stern-rec}}{=}\;0.875\,\frac{C_{R}\,d}{2A_{cs}}\,N\,\tau\label{c_int}\,,
\end{equation}
where $C_{R}$ is the root chord of the fin, $d$ the diameter of
the body tube, $A_{cs}$ again its cross-sectional area, and $N$
is the number of fins. The more accurate drag calculation given in
\cite{Aerodynamic-Drag-Model-Rockets} also includes drag from launch
lugs. The examples given in \cite[p.\ 44, 49]{Aerodynamic-Drag-Model-Rockets}
lead to a small launch lugs drag coefficient of $0.02$ to $0.03$. 
For our rough estimate  we incorporate
this kind of drag by adding $c_{\text{lau}}=0.03$.

\subsection{\label{subsec:Total-drag-coefficient}Total drag coefficient}

Collecting the above discussion, we receive a base value for the
drag coefficient of the model rocket by summation of $c_{\text{nose}}$,
$c_{\text{tube}}$, $c_{\text{base}}$, $c_{\text{fin}}$, $c_{\text{int}}$
and $c_{\text{lau}}$, see  \eqref{c_nose-tube}, \eqref{c_base},
\eqref{cfin-stern-rec}, \eqref{c_fin}, and \eqref{c_int}. However, we have
not considered the surface texture of the rocket so far, ``\emph{roughness}
\emph{can cause drag to increase by about $25$ percent}'', c.f.
\cite[p. 43]{Aerodynamic-Drag-Model-Rockets}. Since our rocket's nose
consists of half a tennis ball, fins and launch lugs are fixed with
hot glue and tape, we presuppose that the surface might be rather
rough. This leads to a significant uncertainty which we incorporate
as an error in the following. Finally, we receive a more or less adequate
formula for the drag coefficient of our D.I.Y. rocket by 
\begin{equation}
c_{D}  =\left[\mathcal{C}_{l}+\frac{0.029}{\sqrt{\mathcal{C}_{l}}}+\frac{0.875\tau}{A_{cs}}\left(A_{\text{fin}}+\frac{C_{R}\,d}{2}\right)N+0.03\right]\cdot\left(1.125\pm0.125\right)\label{cD-Formel}
\end{equation}
with $\mathrm{\mathcal{C}_{l}}$ from \eqref{c_nose-tube}. Beside
the ingredients that enter into $\mathrm{\mathcal{C}_{l}}$, $\tau$
is the thickness to chord ratio of the fins, $C_{R}$ the root chord
of the fin, and $N$ the number of the fins. The skin friction coefficient
$c_{f}$ depends on the kind of air flow and varies greatly with the
Reynolds number, see right side of Figure \ref{fig:Zero-lift-fin-and-skin-friction}.
As $c_{f}$ depends on the rocket's speed it would be the best
to include the skin friction into the system of differential equations
\eqref{DGL-SYSTEM-V-v-h}. However, our estimate of the total drag
coefficient seems to be too vague to justify this approach. Therefore,
we decided to use a constant value for the skin friction coefficient
in the following calculations.

\subsection{An estimate of the drag coefficient of our D.I.Y.\ rocket}

We get an estimate of the drag coefficient by \eqref{cD-Formel}.
The maximum speed of our rocket is about $17\frac{{m}}{{s}}$.
The characteristic length of our rocket is $0.35\,{m}$.
Hence, a kinematic air viscosity of $1,5\cdot10^{-5}\frac{m^{2}}{{s}}$
(at $15^{\circ}C$) leads to a maximum Reynolds number of about
$4\cdot10^{5}$. Based on Prandtl's law \cite[eq.\ 7.43]{White-Fluid-Mechanics}, a mean value for the skin friction coefficient on this scale is given by 
\[
\bar{c}_{f}=\frac{1}{4\cdot10^{5}}\int_{0}^{4\cdot10^{5}}\frac{0.027}{Re_{x}^{1/7}}\,dRe_{x}\approx0.005.
\]
Indeed, this value is in accordance with the skin friction coefficient
at a Reynolds number of $4\cdot10^{5}$ given in \cite[fig.\ A-1]{Aerodynamic-Drag-Model-Rockets}.
Our rocket has a diameter of $8\,{cm}$ which leads to a cross-sectional
area of $A_{cs}\approx0.005\,m^{2}$. The rocket's surface area
(taken as a cylinder with patched-up half of a tennis
ball\footnote{Drag coefficients of a non spinning tennis balls have been measured
to $0.65\pm0.05$, see \cite{tennis-ball}.}) amounts to $A_{ws}\approx0.1\,m^{2}$. The rocket has $N=3$
fins. Each has a root chord of $7.5\,{cm}$ and a fin area of
$A_{\text{fin}}\approx0.0035\,m^{2}$. The fin-thickness-to-chord ratio is about $0.08$. The drag coefficient  \eqref{cD-Formel} is 
\[
c_{D}=0.57\pm0.06,
\]
for a {\tt wxMaxima} source code see again \href{https://github.com/tguent/code}{https://github.com/tguent/code}.

\subsection{Wind tunnel experiment at TU Dortmund University}

Prof.\ Dr.\ Andreas Br\"ummer of the chair of Fluidics at TU Dortmund
University kindly provided us the opportunity to check our estimate
of the drag coefficient in a wind tunnel. The drag force was measured
with a dynamometer with an error of $0.05\,{N}$. We
always had to do an additional measurement for the calibration. Thus,
in the worst case scenario our error adds up to $\Delta F_{D}=0.1\,{N}$.
The wind speed was given to the first decimal. Hence, we use
an error of $\Delta v=0.05\frac{{m}}{{s}}$ in the following
considerations. The drag coefficient $c_{D}$ is related to drag force
$F_{D}$ and wind speed $v$ by $c_{D}=\frac{2F_{D}}{\rho_{\text{air}}A_{R}v^{2}}$.
Therefore, the error of the drag coefficient can be calculated by 
$
\Delta c_{D}
= c_{D}\left(\frac{\Delta F_{D}}{F_{D}}+2\frac{\Delta v}{v}\right)
$. 

\begin{table}[htb]\centering
\begin{tabular}{c|c|c}
Wind speed $v/\frac{m}{s}$			&Drag Force $F_D/N$			& Drag coefficient $c_D$ 	\\\hline
$9.6\pm0.05 $	&$0.20\pm0.1$	&$0.7\pm0.4$\\
$11.3\pm0.05$	&$0.20\pm0.1$	&$0.5\pm0.3$\\	
$12.8\pm0.05$	&$0.25\pm0.1$	&$0.5\pm0.2$\\
$13.8\pm0.05$	&$0.30\pm0.1$	&$0.5\pm0.2$\\
$14.6\pm0.05$	&$0.40\pm0.1$	&$0.6\pm0.2$\\
$15.9\pm0.05$	&$0.40\pm0.1$	&$0.5\pm0.1$\\
$16.8\pm0.05$	&$0.45\pm0.1$	&$0.5\pm0.1$\\
$18.3\pm0.05$	&$0.60\pm0.1$	&$0.6\pm0.2$\\
$18.9\pm0.05$	&$0.6\pm0.10$	&$0.6\pm0.1$\\
$19.8\pm0.05$	&$0.65\pm0.1$	&$0.6\pm0.1$\\
$20.0\pm0.05$	&$0.70\pm0.1$	&$0.6\pm0.1$
\end{tabular}

\caption{The results from the wind tunnel experiment}\label{table1}
\end{table}

\begin{figure}[htb]\centering
\includegraphics[width=0.75\textwidth]{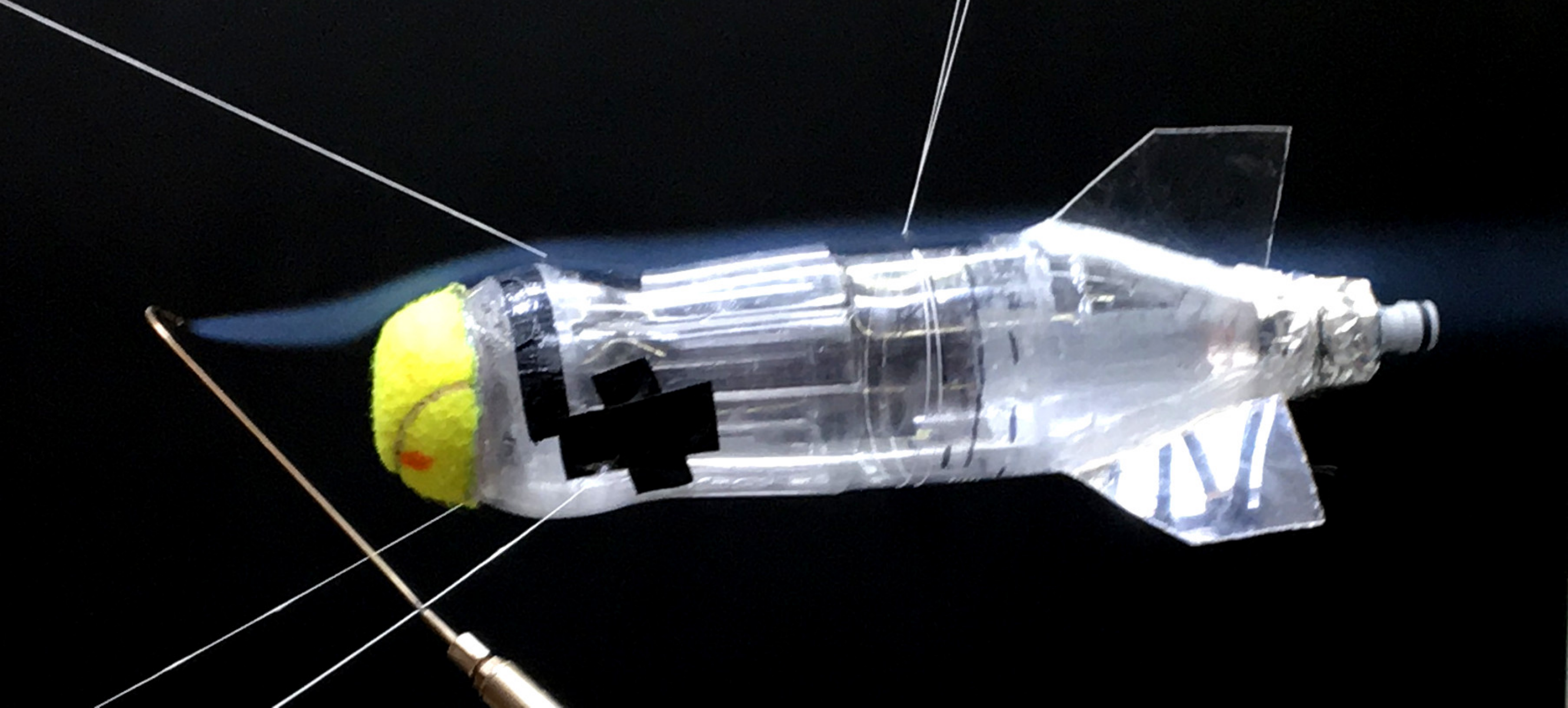}
\caption{\label{fig:Streamline-Flow-Pattern}Stream line flow pattern made visible
by smoke in the wind tunnel at the Department of Fluidics at TU Dortmund
University.}
\end{figure}

From the wind tunnel experiment we receive a drag coefficient of 
\[
c_{D}=0.6\pm0.2\,,
\]
see Table \ref{table1}. In fact, this is a remarkable low value for our D.I.Y.\ rocket. However, from 
Figure \ref{fig:Streamline-Flow-Pattern} we see that the stream line flow
pattern suggests very good aerodynamics properties.

\section{The rocket launch experiment}

The rocket launch was repeated with different initial values for pressure, amount of water, and temperature (of water). 
The model rocket which was used for the experiment has weight $143\,g$. 
It was constructed from a bottle with a volume of $1000\,ml$. 
At launch the rocket was filled with an amount of water $V_l$ and air $V_0$ at pressure $p_0$ and temperature $T$. 
The difference to atmospheric pressure $p_{a}\approx1\,{bar}$ is  $p_{0}-p_{a}$. 
The pressure $p_0$ was measured with an error of $\Delta p = 0.05\,bar$. 
The error is based on the manometer's scale. 
The rocket has a radius of $40\,mm$. 
We drew a scale for the water level which is legible up to about $\pm2\,mm$. 
Within the relevant scope the rocket is cylindric. 
Accordingly, the initial amount of water $V_l$ -- and therewith the air volume $V_0$ -- is determined up to an error of $\Delta V = 10\,ml$. 
Due to the precision of temperature measurements the error of temperature doesn't affect the results. 
Determining the rocket's altitude $h_{\text{mes}}$ makes up the major part in measuring inaccuracy within our experiment. 
We placed a measuring rod beside the launching device. 
The whole flight phase was filmed and we provide links to the corresponding videos in Table \ref{table4}. 
By evaluating the videos we receive the rocket's maximum altitude by comparing to the length of the measuring rod. 
The extrapolation leads to an error of $\Delta h_{\text{med}}=0.5\, m$. 
The measured altitudes can be confirmed by the given video links.

It is the aim of this section to compare the experimental data with the theoretical results, see Table \ref{table3}. 
We use the mean values of the experimental data for the theoretical results. 
These are calculated as described in Section \ref{subsec:Thrust-phase-including}. 
Thereby, $h_{\text{calc}}$ is calculated by using the system of differential equations \eqref{DGL-SYSTEM-V-v-h} for the thrust phase.
We get $h_{\text{est}}$ by calculating the part of the altitude that is obtained from the water thrust phase from  \eqref{eq:3}. 
This can be done with a simple graphing calculator. 
The rocket's drag coefficient was set to $c_D=0.6$. For the air thrust efficiency factor we use $\mu=3$ in \eqref{eta-efficiency-factor}.

A {\tt wxMaxima} source code for the calculations from Table \ref{table3} is available at
\href{https://github.com/tguent/code}{https://github.com/tguent/code}.

\begin{table}[htb]\centering
\begin{tabular}{c|c|c|c|c|c}
 & $T/{}^\circ C$ & $V_{l}/ml$ & $V_{0}/ml$ & $p_{0}/bar$ & $h_{\text{mes}}/m$ \\\hline 
1&$17$ & $350\pm10$ & $650\pm10$ & $3.5\pm0.05$ &  $17.5\pm0.5\,{m}$ \\
2&$17$ & $325\pm10$ & $675\pm10$ & $3.5\pm0.05$ &  $17.1\pm0.5\,{m}$ \\
3&$17$ & $255\pm10$ & $745\pm10$ & $3.0\pm0.05$ &  $14.1\pm0.5\,{m}$ \\
4&$17$ & $255\pm10$ & $745\pm10$ & $3.5\pm0.05$ &  $17.4\pm0.5\,{m}$ \\
5&$20$ & $0$ & $1000$ & $2.5\pm0.05$ & $>2\,{m}$ \\
6&$35$ & $250\pm10$ & $750\pm10$ & $2.5\pm0.05$ &  $7.6\pm0.5\,{m}$ 
\end{tabular}

\caption{The results from the rocket experiment}
\label{table2}\end{table}

\begin{table}[htb]\centering
\begin{tabular}{c|c|c|c|c|c|c}
 & $T/{}^\circ C$ & $V_{l}/ml$ & $V_{0}/ml$ & $p_{0}/bar$ & $h_{\text{calc}}/m$ & $h_{\text{est}}/m$  \\
\hline 
1&$17$ & $350$ & $650$ & $3.5$ & $17.2\,{m}$ & $17.5\,{m}$ \\
2&$17$ & $325$ & $675$ & $3.5$ & $17.4\,{m}$ & $17.6\,{m}$ \\ 
3&$17$ & $255$ & $745$ & $3.0$ & $13.3\,{m}$ & $13.4\,{m}$ \\
4&$17$ & $255$ & $745$ & $3.5$ & $17.2\,{m}$ & $17.3\,{m}$ \\
5&$20$ & $0$ &  $1000$ & $2.5$& $2.4\,{m}$ & $2.4\,{m}$ \\
6&$35$ & $250$ & $750$ & $2.5$ & $9.5\,{m}$ & $9.5\,{m}$ 
\end{tabular}

\caption{Theoretical predictions}
\label{table3}\end{table}

\begin{table}[htb]\centering
\begin{tabular}{c|c}
& \text{video link}\\\hline
1&\href{https://youtu.be/YXGb75eOqbI}{https://youtu.be/YXGb75eOqbI}\\
2& \href{https://youtu.be/9XY_6iSHXPE}{https://youtu.be/9XY\_{}6iSHXPE}\\
3&\href{https://youtu.be/3l5JJHO0Tp8}{https://youtu.be/3l5JJHO0Tp8}\\
4& \href{https://youtu.be/RUkJwDCq8uU}{https://youtu.be/RUkJwDCq8uU}\\
5& \href{https://youtu.be/MifjHZH3Z7Y}{https://youtu.be/MifjHZH3Z7Y}\\
6& \href{https://youtu.be/8MRufGWCUjM}{https://youtu.be/8MRufGWCUjM}
\end{tabular}

\caption{Video links}
\label{table4}\end{table}

\section{Conclusion}

The method introduced in this paper yields accurate theoretical predictions. 
The predictions in launches 1, 2 and 4 from Table \ref{table4} lie within the experimental error, see Table \ref{table2}. 
Launch 5 was hard to evaluate because the rocket crashed into the ceiling. 
Indeed, the main goal of launch 5 was to show that there is significant thrust
if the rocket is filled with pressured air only. 
Launch 3 yielded an altitude of $13.6\,m$ at the lower bound 
which differs from the predicted value of $13.4\,m$. 
The question comes up whether the error in altitude has to be enlarged. 
On the one hand, the camera position has to be far enough from the experiment 
that the elevation angle remains small. 
On the other hand, huge distance leads to more blurred pictures. 
It stands to reason that one has to find a more suitable method to evaluate the maximum altitude. 
In case of launch 6, theoretical and experimental results don't coincide. 
Indeed, we launched the rocket several times without getting exploitable data because the rocket didn't lift off correctly. In some cases the rocket stuck too long to the launching device due to friction. 
In other cases the rocket lurched through the air. 
Furthermore, our model D.I.Y. rocket began to leak after a huge amount of experiments. 

Putting it all together, our method is suitable to predict the altitude in case that the rocket lifts off perfectly. 
But even in this case the water rocket physics represents a highly chaotic system. 
On this basis, the simple estimation proposed in this paper yields amazingly good results.


\begin{thebibliography}{10}
\bibitem{Abramowitz} 
M.~Abramowitz, I.~A.~Stegun: 
\emph{Handbook of mathematical functions with formulas, graphs, and mathematical tables}. 
Dover Publications, 1965

\bibitem{DragTest2009}
R.~Barrio-Perotti, E.~Blanco-Marigorta, K.~Arguelles-Diaz, J.~Fernandez-Oro: 
Experimental Evaluation of the Drag Coefficient of Water Rockets by a Simple Free-Fall Test. 
{\em European Journal of Physics} {\bf 30} no.~5, 1039-1048 (2009)

\bibitem{Barrio2010}
R.~Barrio-Perotti, E.~Blanco-Marigorta, J.~Fernández-Francos, M.~Galdo-Vega: 
Theoretical and experimental analysis of the physics of water rockets.
{\em European Journal of Physics} {\bf 31} no.~5, 1131-1147 (2010)

\bibitem{Bronstein} 
I.~N.~Bronstein: 
\emph{Taschenbuch der Mathematik}.
B.~G.~Teubner Stuttgart-Leipzig, 1996

\bibitem{Buck}
A.~L.~Buck: 
New equations for computing vapor pressure and enhancement factor. 
{\em J.~Appl.~Meteorol.} {\bf 20} no.~12, 1527-1532 (1981)

\bibitem{Campbell2017}
T.~A.~Campbell, M.~Okutsu: 
Model Rocket Project for Aerospace Engineering Course: Trajectory Simulation and Propellant Analysis. 
Preprint 2017, 
\href{https://arxiv.org/pdf/1708.01970.pdf}{arXiv:1708.01970 [physics.ed-ph]}

\bibitem{Aerodynamics-Clancy}
L.~J.~Clancy: 
\emph{Aerodynamics}. 
John Wiley \& Sons, 1975 

\bibitem{Clifford2006}
M.~Clifford (ed.): 
\emph{An Introduction to Mechanical Engineering}. 
CRC Press Taylor \& Francis Group, 2006 (Part 1), 
Hodder Education, An Hachette UK Company, 2010 (Part 2)

\bibitem{Finney1999}
G.~A.~Finney: 
Analysis of a water-propelled rocket: A problem in honors physics.
{\em American Journal of Physics} {\bf  68}  no.~3, 223-227 (2000)

\bibitem{Gommes2010}
C.J.\ Gommes: A more thorough analysis of water rockets: Moist adiabats, transient flows and inertial forces in a soda bottle. 
{\em American Journal of Physics} {\bf 78} no.~3, 236-243 (2010)  
\href{https://doi.org/10.1119/1.3257702}{DOI: 10.1119/1.3257702}

\bibitem{Aerodynamic-Drag-Model-Rockets}
G.~M.~Gregorek, \emph{Aerodynamic Drag of Model Rockets}. 
Estes Industries, Penrose, CO, 1970

\bibitem{Manual-Atmosphere}
International Civil Aviation Organization:
\emph{Manual of the ICAO Standard Atmosphere}. 
Doc 7488/3, 3rd ed.\, 1993

\bibitem{tennis-ball}
R.~Mehta, F.~Alam, A.~Subic: 
Review of tennis ball aerodynamics.
{\em  Sports Technology} {\bf1} no.~1, 7-16 (2008)   
\href{https://doi.org/10.1080/19346182.2008.9648446}{DOI: 10.1080/19346182.2008.9648446}

\bibitem{soda-bottle}
D.~Kagan, L.\ Buchholtz, L.\ Klein: 
Soda-bottle water rockets.
{\em Phys.~Teach.} {\bf 33} 150, 1995

\bibitem{Raumfahrtsysteme}
E.~Messerschmid und S.~Fasoulas: 
Die Ziolkowsky-Raketengleichung.
Chapter 2 of {\em Raumfahrtsysteme}.  
Springer Vieweg, Berlin, Heidelberg, 2017,
\href{https://doi.org/10.1007/978-3-662-49638-1_2}{DOI: 10.1007/978-3-662-49638-1\_{}2}

\bibitem{Moran1984}
J.~Moran, 
\emph{An Introduction to Theoretical and Computational Aerodynamics}.
John Wiley \& Sons, New York, 1984

\bibitem{Nelson1976}
R.~A.~Nelson, M.~E.~Wilson: 
Mathematical analysis of a model rocket trajectory Part I: The powered phase. 
{\em Phys.~Teach.} {\bf 14} no.~3, 150-161 (1976)

\bibitem{Prusa2000}
J.~M.~Prusa: Hydrodynamics of a Water Rocket. 
{\em Siam Rev.} {\bf 42} no.~4, 719-726 (2000)

\bibitem{Air-Expansion-2013}
A.~Romanelli, I.~Bove, F.~G.~Madina: 
Air expansion in the water rocket. 
{\em American Journal of Physics} {\bf  81} no.\10, 762-766 (2013) 
\href{https://doi.org/10.1119/1.4811116}{DOI:10.1119/1.4811116}

\bibitem{SimpleDragTest}
Simple Drag Tests for Water Rockets - seeds2lrn.com. 
\href{http://fliphtml5.com/rftx/obtf}{http://fliphtml5.com/rftx/obtf}
(visited 05.03.2019)

\bibitem{Slater1966}
L.~J.~Slater: 
\emph{Generalized Hyperbolic Functions}.
Cambridge University Press, 1966

\bibitem{St=0000F6cker}
H.~St\"ocker: 
\emph{Taschenbuch der Physik}.
Verlag Harri Deutsch, 1998

\bibitem{Tsiolkovsky}
K.~E.~Tsiolkovsky: 
The Exploration of Cosmic Space by Means of Reaction Devices (in Russian). 
{\em The Science Review } {\bf 5} (1903)

\bibitem{MAXIMA}
A.~Vodopivec: 
wxMaxima 18.02.0. 
\href{http://andrejv.github.io/wxmaxima/}{http://andrejv.github.io/wxmaxima/}

\bibitem{White-Fluid-Mechanics}
F.~M.~White: 
\emph{Fluid Mechanics}.
McGraw-Hill Education Ltd, 7th ed., 2011
\end{thebibliography}
\end{document}